\documentclass[12pt]{article}
\pdfoutput=1 %pdflatex
\usepackage{amsmath,amssymb,amsfonts,amsthm}
\usepackage{float,graphicx, multirow, setspace}
\usepackage{psfrag,epsf}
\usepackage{enumerate}
\usepackage{subfiles}
\usepackage{placeins}
\usepackage{xurl}
\usepackage{xr,xr-hyper,hyperref}
\usepackage{algpseudocode}
\usepackage{subfig}
\usepackage{algorithm}
\usepackage{makecell}
\usepackage[scr=boondox]{mathalfa}
\usepackage{xspace}

\usepackage{afterpage}

 \newcommand{\vmarket}{{ v_{\text{market}}}}

 \newcommand{\ybust}{{ y_{\text{bust}}}}
 
 \newcommand{\cost}{{ \mathrm{cost} }}

 \newcommand{\pos}{{ \mathsf{pos}}}
 \newcommand{\qb}{{ \mathsf{qb}}}
 \newcommand{\notqb}{{ \mathsf{not \ qb}}}
 
 \newcommand{\logistic}{{ \mathrm{logistic}}}
  
  \newcommand{\bp}{{ \mathrm{bp}}}
  \newcommand{\Unif}{{ \mathrm{Unif} }}
  \newcommand{\Beta}{{ \mathrm{Beta} }}
  \newcommand{\Bernoulli}{{ \mathrm{Bernoulli}}}

 \newcommand{\bE}{{\mathbb E}}
 \newcommand{\Var}{\mathrm{Var}}
 \newcommand{\sd}{{ \mathrm{sd}}}
 
 \newcommand{\shapeone}{\mathrm{shape}_1}
 \newcommand{\shapetwo}{\mathrm{shape}_2}
 \newcommand{\bP}{{\mathbb P}}

 \newcommand{\R}{{\mathsf R}}

\newcommand{\blind}{0}

\usepackage[round]{natbib}

\usepackage[dvipsnames]{xcolor}
\usepackage[margin=1in]{geometry}

\usepackage{fullpage}
\usepackage{parskip}

\numberwithin{equation}{section}
\definecolor{SkyBlue}{RGB}{14, 118, 188}
\definecolor{BrightRed}{RGB}{223,82, 78}
\hypersetup{pdfborder = {0 0 0.5 [3 3]}, colorlinks = true, linkcolor = BrightRed, citecolor = SkyBlue, filecolor = BrightRed}
\title{
  The Loser's Curse and the Critical Role of the Utility Function
}

\begin{document}

\def\spacingset#1{\renewcommand{\baselinestretch}%
{#1}\small\normalsize} \spacingset{1}

\if1\blind
{
  \author{Authors blinded for review}
} \fi
\if0\blind
{
  \author{Ryan S. Brill\thanks{Graduate Group in Applied Mathematics and Computational Science, University of Pennsylvania. Correspondence to: ryguy123@sas.upenn.edu} \ and Abraham J. Wyner\thanks{Dept.~of Statistics and Data Science, The Wharton School, University of Pennsylvania}}
} \fi

\maketitle

\bigskip
\begin{abstract}
A longstanding question in the judgment and decision making literature is whether experts, even in high-stakes environments, exhibit the same cognitive biases observed in controlled experiments with inexperienced participants. Massey and Thaler (2013) claim to have found an example of bias and irrationality in expert decision making: general managers' behavior in the National Football League draft pick trade market. They argue that general managers systematically overvalue top draft picks, which generate less surplus value on average than later first-round picks, a phenomenon known as the loser's curse. Their conclusion hinges on the assumption that general managers should use expected surplus value as their utility function for evaluating draft picks. This assumption, however, is neither explicitly justified nor necessarily aligned with the strategic complexities of constructing a National Football League roster. In this paper, we challenge their framework by considering alternative utility functions, particularly those that emphasize the acquisition of transformational players––those capable of dramatically increasing a team's chances of winning the Super Bowl. Under a decision rule that prioritizes the probability of acquiring elite players, which we construct from a novel Bayesian hierarchical Beta regression model, general managers' draft trade behavior appears rational rather than systematically flawed. More broadly, our findings highlight the critical role of carefully specifying a utility function when evaluating the quality of decisions.
\end{abstract}

\noindent%

\newpage
\spacingset{1.45} % DON'T change the spacing!

% %%%%%%%%%%%%%%%%%%%%%%%%%%%%%
% \documentclass[12pt]{article}
% \usepackage{amsmath,amsfonts,amssymb}
% \usepackage{graphicx,psfrag,epsf}
% \usepackage{enumerate}
% \usepackage[round]{natbib}
% \usepackage{url} 
% \usepackage[dvipsnames]{xcolor}
% \input{header}
% % \doublespace
% \onehalfspacing
% \bibliographystyle{apalike}
% % DON'T change margins - should be 1 inch all around.
% \addtolength{\oddsidemargin}{-.5in}%
% \addtolength{\evensidemargin}{-.5in}%
% \addtolength{\textwidth}{1in}%
% \addtolength{\textheight}{1.3in}%
% \addtolength{\topmargin}{-.8in}%

% \usepackage{fullpage}
% \usepackage{parskip}
% % \usepackage{mathptmx}
% % \newtheorem{theorem}{Theorem}
% % \numberwithin{equation}{section}
% \interfootnotelinepenalty=10000
% \definecolor{SkyBlue}{RGB}{14, 118, 188}
% \definecolor{BrightRed}{RGB}{223,82, 78}
% \hypersetup{pdfborder = {0 0 0.5 [3 3]}, colorlinks = true, linkcolor = BrightRed, citecolor = SkyBlue, filecolor = BrightRed}
% %%%%%%%%%%%%%%%%%%%%%%%%%%%%%
% \begin{document}

% % \begin{abstract}
% % \input{abstract}
% % \end{abstract}

%%%%%%%%%%%%%%%%%%%%%%%%%%%%%%%%%%%%%%%%%%%%%%%%%%%%%%%%%%%%%%%%%%%%%%%%%%%%%%%%%%%%%%%%
%%%%%%%%%%%%%%%%%%%%%%%%%%%%%%%%%%%%%%%%%%%%%%%%%%%%%%%%%%%%%%%%%%%%%%%%%%%%%%%%%%%%%%%%
%%%%%%%%%%%%%%%%%%%%%%%%%%%%%%%%%%%%%%%%%%%%%%%%%%%%%%%%%%%%%%%%%%%%%%%%%%%%%%%%%%%%%%%%
\section{Introduction}\label{sec:intro}

A longstanding question in the judgment and decision making literature is whether experts, when making high-stakes decisions, exhibit the same cognitive biases and irrational behaviors documented in studies involving inexperienced participants in low-stakes environments.
Critics of this literature argue that biases may be overstated, as most studies do not analyze decision making in real-world, high-stakes settings.
\citet{MasseyThaler2013} claim to have found an exception, applying behavioral economic principles to a context––the National Football League (NFL) draft––where decision makers, namely NFL general managers, are highly experienced, financially incentivized, and have access to extensive data and learning opportunities.

Massey and Thaler's study has had a significant impact not just in sports analytics but in the broader decision making literature, reinforcing the notion that systematic biases can persist even in expert settings.
They argue that general managers systematically overpay for top draft picks despite evidence that these selections generate less ``surplus value''––the difference between the value of a player’s performance and his cost––than later first-round picks, on average.
This so-called ``loser’s curse,'' widely cited as evidence of irrational decision making, is the subject of their ultimate striking assertion: ``top draft picks are significantly overvalued in a manner that is inconsistent with rational expectations and efficient markets, and consistent with psychological research'' \citep{MasseyThaler2013}.

In this paper, we challenge that assertion.
Massey and Thaler’s conclusion critically depends on a single questionable assumption, namely that general managers should maximize expected surplus value.
We argue that this assumption is overly simplistic and does not necessarily capture the strategic objectives of NFL general managers.
Winning in the NFL may depend less on accumulating surplus value and more on acquiring transformational players––those capable of fundamentally altering a team's trajectory. 
If general managers instead prioritize maximizing the probability of drafting elite players, their decisions—–while appearing suboptimal under Massey and Thaler’s framework––may in fact be entirely rational.

In other words, we question whether expected surplus value is the proper ``rational'' criterion for evaluating risky prospects.
It is well known that individuals play the lottery and prior research has shown that individuals sometimes have a preference for positively skewed payoffs (e.g. \citet{Barberis}). 
However, a preference for positively skewed payoffs is often viewed as irrational in the context of traditional economic theories, which assume that individuals make decisions based on maximizing expected utility or wealth. 
The present research, in contrast, highlights an important decision context where, under reasonable assumptions, a preference for positively skewed payoffs may be perfectly rational.

This perspective provides a more nuanced view of decision making in high-stakes environments. 
Perhaps irrationality and cognitive biases are not as pervasive as Massey and Thaler, along with the broader judgement and decision making literature, assert. 
Our findings underscore the importance of specifying a decision-maker's utility function before labeling behavior as irrational. 
More broadly, we highlight the risks of applying oversimplified economic models to complex, domain-specific decision-making contexts without considering alternative utility functions.

Below, we begin with an exposition of the setting of our case study––the NFL draft––and then give a more detailed overview of Massey and Thaler's analysis and our rebuttal.

%%%%%%%%%%%%%%%%%%%%%%%%%%%%%%%%%%%%%%%%%%%%%%%%%%%%%%%%%%%%%%%%%%%%%%%%%%%%%
\subsection{Overview}\label{sec:overview}

Players entering the NFL are allocated to 32 teams through an annual draft.
The draft consists of seven rounds, in which teams take turns selecting players in reverse order of the previous season’s win-loss records.
The team with the worst record picks first, while the reigning Super Bowl champion selects last (with some minor exceptions). 
There are up to $256$ total draft picks because the NFL adds up to $32$ additional ``compensatory free agent'' draft picks––creating a hidden eigth round––that allow teams to replace players who they lost in free agency.\footnote{
 \url{https://operations.nfl.com/journey-to-the-nfl/the-nfl-draft/the-rules-of-the-draft}
}
Trading draft picks, involving the exchange of current and future draft selections as well as active players, is a fundamental component of constructing an NFL roster.
Teams typically engage in these trades to secure specific players or improve positional depth. 
For example, in 2013 the Miami Dolphins traded the $3^{rd}$ overall pick to the Oakland Raiders in exchange for the $12^{th}$ and $42^{nd}$ overall picks.\footnote{
    \url{https://bleacherreport.com/articles/1618345-miami-dolphins-select-dion-jordan-with-no-3-pick-in-trade-with-oakland-raiders}
}
Trades like this raise fundamental questions about the valuation of draft picks.
How did the Dolphins and Raiders determine that their trade was mutually beneficial?
More broadly, if a team seeks to acquire a specific draft pick, what constitutes an appropriate offer?
If presented with a trade proposal involving multiple draft picks, how should a team evaluate its value and decide whether to accept or reject the exchange?
A rigorous analysis of draft pick valuation is essential to understanding the trade dynamics that shape NFL roster management.

For decades, analysts and researchers have used \textit{draft position value curves} to quantify the value of draft positions and assess the fairness of trades involving draft picks.
Draft position value curves assign a numerical point value to each draft position $1, ..., 256$ and are typically additive: the value of a bundle of picks is the sum of the values of each individual pick.
A trade is considered fair if the sum of the values on both sides is equal.
In the 1990s, at the request of Dallas Cowboys head coach Jimmy Johnson, vice president Mike McCoy developed the first draft position value curve\footnote{
    \url{https://www.the33rdteam.com/jimmy-johnsons-draft-trade-value-chart-set-standard-but-clearly-outdated/}
} 
based on intuition and an analysis of past trades.
This curve assigns the highest value to the first overall pick and follows a steep, convex decline.
For instance, in the 2013 draft trade between the Dolphins and the Raiders, Johnson’s curve suggests the Dolphins' draft capital increased by 31\%.
The Raiders traded a single pick worth 2,200 points in exchange for two picks collectively worth 1,680 points (1,200 and 480, respectively).
Although Johnson’s chart is ad hoc and lacks formal rigor, it remained widely used by general managers for decades, and some teams continue to rely on it today.

% More recently, in their seminal work, \citet{MasseyThaler2013} developed the first rigorous data-driven draft position value curves, which are visualized in Figure~\ref{fig:MasseyThalerFindings}.

\begin{figure}[hbt!]
    \centering{}
    \subfloat[\centering ]{
        {\includegraphics[width=0.5\textwidth]{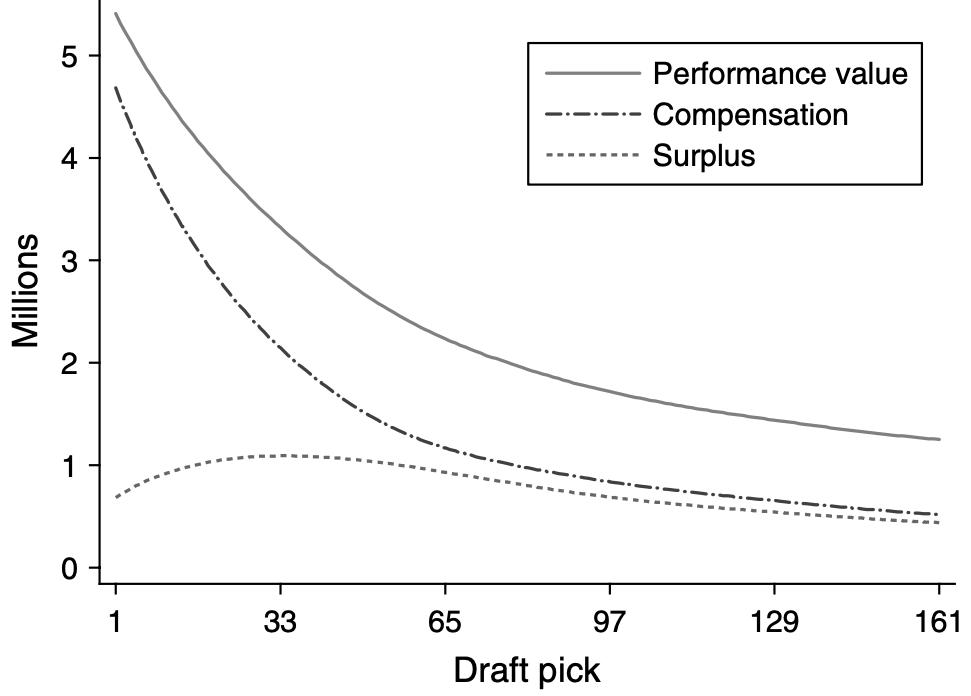}}
        \label{fig:MasseyThalerFinding1}
    }
    % \qquad
    \subfloat[\centering ]{
        {\includegraphics[width=0.5\textwidth]{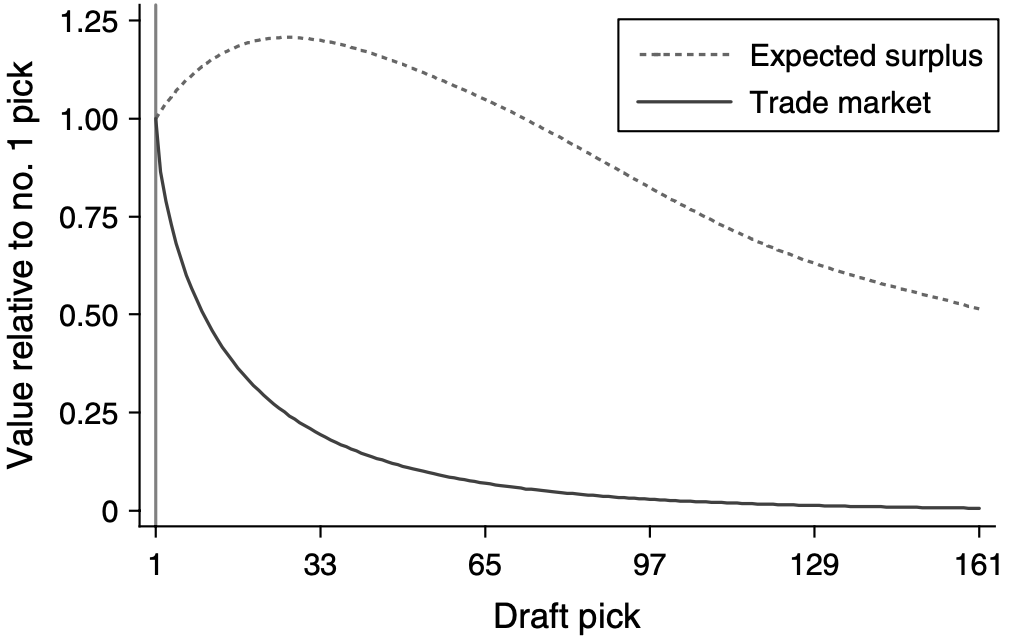}}
        \label{fig:MasseyThalerFinding2}
    }
    \caption{
    Figure 3 from \citet{MasseyThaler2013}, which summarizes their results.
    In (a) they visualize expected performance value (solid line), first contract compensation (dot-dashed line), and expected surplus value (dashed line) versus draft position, expressed on a salary-equivalent scale.
    In (b) they visualize expected surplus value (dashed line) and the trade market curve (solid line) versus draft position, expressed relative to the value of the first pick.
    }
    \label{fig:MasseyThalerFindings}
\end{figure}

More recently, \citet{MasseyThaler2013} developed less ad hoc draft position value curves, visualized in Figure~\ref{fig:MasseyThalerFindings}. 
The solid line in Figure~\ref{fig:MasseyThalerFinding1} denotes expected performance value, which quantifies the expected on-field performance of players drafted at a given position, expressed on a salary-equivalent scale.
The dot-dashed line reflects first contract compensation, which denotes the cost of drafted players during their first contract.
A player's first contract compensation is essentially deterministic, as it is dictated by draft position.\footnote{Drafted rookies may negotiate certain contract elements, such as performance bonuses or injury guarantees.}
However, this compensation is often much lower than the value of the player’s on-field performance, leading to what is known as first contract surplus value.
The dashed lines in Figures~\ref{fig:MasseyThalerFinding1} and \ref{fig:MasseyThalerFinding2} reflect expected surplus value, which measures the expected difference between a player's first contract performance and his cost.
Finally, the solid line in Figure~\ref{fig:MasseyThalerFinding2} is the trade market curve, a draft position value curve that best fits the dataset of all observed trades involving only draft picks.
This curve reflects the draft value implied by actual trade market behavior.

%%%%%%%%%%%%%%%%%%
%%%%%%%%%%%%%%%%%%

Massey and Thaler asserted two main takeaways from their analysis displayed in Figure~\ref{fig:MasseyThalerFindings}.
First, they claimed to have found a \textit{loser's curse}, visualized in Figure~\ref{fig:MasseyThalerFinding1}.
Teams with the worst win-loss records in the previous season—the losers—receive the top draft picks, which produce higher performance on average since expected performance value declines monotonically and convexly throughout the draft.
However, these teams are cursed because top draft picks generate less surplus value on average than mid-to-late first-round picks, which typically belong to better teams.
Expected surplus value (i.e., expected performance above cost) increases throughout the early first round, peaking in the middle of the round.
Second, Massey and Thaler asserted that, if one adopts the expected surplus value utility function, the draft pick trade market is highly inefficient, visualized in Figure~\ref{fig:MasseyThalerFinding2}.
The trade market curve is much steeper than the expected surplus value curve and differs in shape.
Thus, teams consistently overpay to trade up in the draft, investing excessive draft capital to select ``top draft picks [that] are significantly overvalued'' \citep{MasseyThaler2013}.
In contrast, trading down and accumulating additional draft picks is an effective strategy to maximize surplus value.
For example, consider the 2013 draft trade between the Dolphins and the Raiders.
Using expected surplus value,\footnote{
    This calculation is based on our expected surplus value curve, visualized in Figure~\ref{fig:plot_Massey_Thaler_replication_1} in Section~\ref{sec:trad_value_chart}, as the exact values from Massey and Thaler’s curve are unavailable.
} 
the Raiders' draft capital increased by 135\%.
The Dolphins traded two picks collectively worth 2.40 points (1.51 and 0.89, respectively) in exchange for a single pick worth 1.02.
Notably, the $12^{th}$ overall pick had a higher expected surplus value than the $3^{rd}$ overall pick (i.e., $1.51 > 1.02$).
Although the Dolphins won the trade according to Jimmy Johnson’s draft value chart, the Raiders accumulated substantially more expected surplus value.

Massey and Thaler's assertions, summarized in Figure~\ref{fig:MasseyThalerFindings}, replicate to recent data (see Figure~\ref{fig:plot_Massey_Thaler_replication_1} in Section~\ref{sec:trad_value_chart}).
% Massey and Thaler's findings, summarized in Figure~\ref{fig:MasseyThalerFindings}, are corroborated by recent data\footnote{
%     That Massey and Thaler's results replicate in recent data was originally noticed as part the development of a forthcoming paper \citeauthor[]{avery}
% } (see Figure~\ref{fig:plot_Massey_Thaler_replication_1} in Section~\ref{sec:trad_value_chart}).
Despite the widespread recognition of their work over an extended period, the draft strategies employed by general managers in the NFL have remained largely unchanged. 
This persistence raises an important question: why has there been no apparent adjustment in decision making? 
One possible explanation is that NFL general managers are resistant to change and continue to engage in suboptimal trade practices due to cognitive biases or misjudgment. 
% One possible explanation––examined in the forthcoming paper \citeauthor[]{avery}––is that NFL general managers are resistant to change and continue to engage in suboptimal trade practices due to cognitive biases or misjudgment. 
However, this perspective prompts further inquiry: are these executives, responsible for managing multi-billion-dollar franchises, fundamentally misguided, or is there a crucial element missing from existing analyses? 
This paper explores the latter hypothesis, proposing an alternative explanation. 
Specifically, we argue that there may be a sound economic rationale underlying observed trade behavior. 
It is plausible that general managers are employing a different, yet entirely rational, utility function that does not rely solely on the expected future surplus value of a player at a given draft position.

Perhaps the overarching strategic objective of NFL teams is to win the Super Bowl––a goal consistently emphasized by team owners, general managers, coaches, and players. As Kansas City Chiefs quarterback Patrick Mahomes succinctly stated, ``the only goal is to win the Super Bowl.''\footnote{\url{
    https://x.com/_MLFootball/status/1826003252460269907?t=JvStTWyM92jEe21LxqOTPw&s=19
}}
Massey and Thaler’s conclusions are predicated on the assumption that general managers, in pursuit of this objective, should adopt expected surplus value as their guiding utility function. 
However, while their analysis relies on this assumption, they do not explicitly acknowledge the primacy of winning the Super Bowl or establish a direct link between maximizing expected surplus value and achieving that goal. 
While it is possible that maximizing expected surplus value contributes to winning, alternative utility functions may yield entirely different strategic conclusions. 
This paper develops that possibility, exploring the idea that Super Bowl success hinges on acquiring transformational players––those capable of profoundly altering a team’s trajectory––rather than optimizing expected surplus value. 
One example is the selection of Patrick Mahomes in the 2017 NFL Draft, which fundamentally elevated the Chiefs from a playoff-caliber team to a perennial Super Bowl contender.\footnote{
    From 2013 to 2017, under coach Andy Reid but without starting quarterback Patrick Mahomes, the Chiefs made the playoffs in four of five seasons but won just one playoff game.
    From 2018 to 2023, the Chiefs had four Super Bowl appearances and won three.
}
Another is Josh Allen, selected in 2018, who elevated the Buffalo Bills from years of ineptitude to one of the league's top teams.\footnote{
    From 2000 to 2017, the Bills made the playoffs once (in 2017). From 2018 to 2024, the Bills made the playoffs six times and made the AFC champsionship game twice.
}
Under a different, but perhaps entirely reasonable, utility function––maximizing the probability of drafting transformational or elite players––decisions that appear suboptimal under Massey and Thaler’s framework may actually be entirely rational.

It is possible that the most effective strategy for constructing a Super Bowl caliber roster is not maximizing expected surplus value but rather valuing draft positions based on the likelihood of acquiring elite talent.
These two objective functions––maximizing expected surplus value versus maximizing the expected number of elite players––differ significantly in their strategic implications. 
While both are based on expected values, the latter considers the probability that a player will produce exceptional, high-impact performance, thereby capturing the importance of \textit{tail} outcomes in shaping team success.

A key insight, overlooked by previous researchers, is how \textit{variance} in performance value changes over the course of the draft.
Just as expected performance value declines convexly throughout the draft (see Figure~\ref{fig:MasseyThalerFinding1}), so does variance (see Figure~\ref{fig:plot_empMeanSd} in Section~\ref{sec:research_hypothesis}).
We hypothesize that general managers are not only trading for expected surplus value but also for variance, which is directly related to the probability of acquiring a transformational player in the far right tail of performance.
For example, consider the 2013 draft trade between the Dolphins and the Raiders.
In this exchange, the Dolphins traded expected surplus value to the Raiders in exchange for variance.
Based on the expected number of elite players,\footnote{
    Here, we define an elite player as one at least as valuable as Jared Goff, represented by the orange curve in Figure~\ref{fig:plot_tail_probs} in Section~\ref{sec:results}.
} the Dolphins’ draft capital increased by 40\%.
The Raiders traded a single pick worth 0.893 expected elite players for two picks collectively worth 0.639 (0.538 and 0.101).
Although the Raiders accumulated substantially more expected surplus value, the Dolphins increased their expected number of elite players.
Further, notice how much less the $42^{nd}$ pick is worth than before––it produces 89\% fewer expected elite players than the $3^{rd}$ pick, but just 13\% less expected surplus value.
The likelihood of acquiring transformational players decays precipitously across the draft.

Hence, in this paper we construct Bayesian hierarchical draft position value curves that reflect the probability of drafting an elite player.
We find that general managers’ behavior in the trade market aligns with a decision rule that prioritizes variance and right-tail outcomes––seeking players with the potential to be game-changing superstars, even if they are riskier investments. 
Quarterbacks in particular follow a more extreme version of this pattern.
Under the lens of drafting elite players, particularly for teams seeking a franchise quarterback, general managers' behavior in the draft pick trade market appears rational rather than systematically flawed.

The remainder of this paper is organized as follows. 
We begin in Section~\ref{sec:measuringPlayerPerformance} by defining the value of an NFL player's performance.
Then, in Section~\ref{sec:trad_value_chart}, we review the traditional approach to constructing draft position value curves, similar to those devised in \citet{MasseyThaler2013}.
In Section~\ref{sec:research_hypothesis} we elaborate on our research hypothesis,  describing how general managers' decision making may be rational under alternative utility functions that prioritize variance and right-tail outcomes over expected surplus value. 
To construct other draft position value curves built from alternative value functions, we estimate the full conditional density of player performance given draft position and player position in Section~\ref{app:estimate_con_density}.
We present our results in Section~\ref{sec:results} and conclude in Section~\ref{sec:discussion}.

%%%%%%%%%%%%%%%%%%%%%%%%%%%%%%%%%%%%%%%%%%%%%%%%%%%%%%%%%%%%%%%%%%%%%%%%%%%%%%%%%%%%%%%%
\section{Measuring player performance}\label{sec:measuringPlayerPerformance}

Since the value of a draft pick is realized by the on-field performance of the player it yields, a crucial first step is to measure the value of a player's performance.
Doing so takes care and consideration.
Each play in American football consists of a complex interplay between 22 players of varying roles interacting in continuous time and space.
This makes it difficult to measure how a player's individual performance impacts the ultimate outcome of a play.
For instance, a linebacker may impact a play simply by occupying space.
He could cause the ball carrier to move in a different direction, and there is no easy way to measure the impact of that action.
There is no industry-standard way of measuring player performance on a given play in American football, particularly in a way that measures players of all positions on the same scale.
Hence, analysts use performance proxies.
Each measure is imperfect and has its own drawbacks.
We describe many of these performance measures in detail in Appendix~\ref{app:performanceValue}.
Throughout this paper, for simplicity, we use one performance outcome variable, detailed below.

In free agency, NFL players who are not bound by a contract can sign a new contract with any team. 
Following Massey and Thaler, we assume the free agent market is relatively efficient and thus reflects the performance value of players.
Therefore, following \citet{BaldwinPosCurves}, in this study we measure performance value by the average salary per year of a contract signed in free agency.
We consider salary as a percentage of the cap––a rule that determines the maximum amount any team can pay its players in a given year––to adjust for inflation.
Over time the cap has increased from \$123 million in 2013 to \$224.8 million in 2023.

Researchers, beginning with \citet{MasseyThaler2013}, have been interested in particular in the value of a player's performance during his first contract.
This is because an NFL player is paid particularly cheaply during his first contract.
A player's first contract is determined by his draft position and can be much lower than the value of his performance, creating a first contract ``surplus value.''
Teams can take advantage of the resulting extra salary cap space by signing good veteran players in free agency.
The drawback of focusing solely on a player's first contract is that the very best players stay with the same team across multiple contracts, providing value across their entire careers.
% For example, Tom Brady played in the NFL for 23 seasons and won seven Super Bowls.
For example, Tom Brady played in the NFL for 23 seasons and won seven Super Bowls, playing almost the entirety of his career with the team that drafted him.
Continuing with the precedent set by \citet{MasseyThaler2013} and subsequent analyses, in this study we consider just first contract performance and surplus value.

Throughout this paper, we measure a player's first contract performance by his second contract's average salary per year as a percentage of the cap.
This is a reflection of the value of a player's first contract performance because he signs his second contract in free agency just after his first contract ends.
Players who don't sign a second contract have zero performance value.
We acknolwedge that free agent annualized cap percentage is an imperfect measure of a player's performance (see Appendix~\ref{app:performanceValue}) and first contract value is an incomplete measure of the total value he provides to his team (as it excludes the rest of his career).
Nonetheless, the focus of this paper is to value draft picks given a performance measure, not to choose the best performance measure.
We leave further analysis of various measures of performance to future work.

%%%%%%%%%%%%%%%%%%%%%%%%%%%%%%%%%%%%%%%%%%%%%%%%%%%%%%%%%%%%%%%%%%%%%%%%%%%%%%%%%
%%%%%%%%%%%%%%%%%%%%%%%%%%%%%%%%%%%%%%%%%%%%%%%%%%%%%%%%%%%%%%%%%%%%%%%%%%%%%%%%%
\section{Traditional draft position value curves}\label{sec:trad_value_chart}

We review the traditional approach to constructing draft position value curves. 
Prior to the draft, we do not know how well a player drafted at draft position $x \in \{1,...,256\}$ will perform in the NFL.
Thus, we think of the performance outcome $Y$ associated with position $x$ as a random variable, denoted by a capital letter.
Recall that, throughout this paper, we let the realization of $Y$ be a player's annualized second contract value as a percentage of the salary cap.
Then, from the dataset of all recent draft picks in NFL history, we want to estimate a de-noised relationship between pick number $x$ and performance outcome $Y$.
Traditional analyses use a standard approach but with different measures of performance, detailed in Appendix~\ref{app:performanceValue} \citep{MasseyThaler2013,StuartDraftChart,OTCDraftChart,PFFDraftChart,BaldwinPosCurves}.
The de-noised relationship they estimate from data is the expected value of $Y$ given $x$.

This traditional approach proceeds as follows.
First, fit the expected performance value curve, or the conditional mean $x \mapsto \bE[Y|x]$.
Second, calculate the cost curve, $x \mapsto \cost(x)$.
Cost is first contract compensation, which is essentially a deterministic function of $x$.
Third, calculate the expected surplus value curve, or the expected value of the difference between performance and cost, $x \mapsto \bE[Y - \cost(x)|x]$.
Finally, normalize each of these curves by dividing by the value of the first pick.
Value is then relative to the first pick, whose value is one.

\begin{figure}[hbt!]
    \centering{}
    \includegraphics[width=\textwidth]{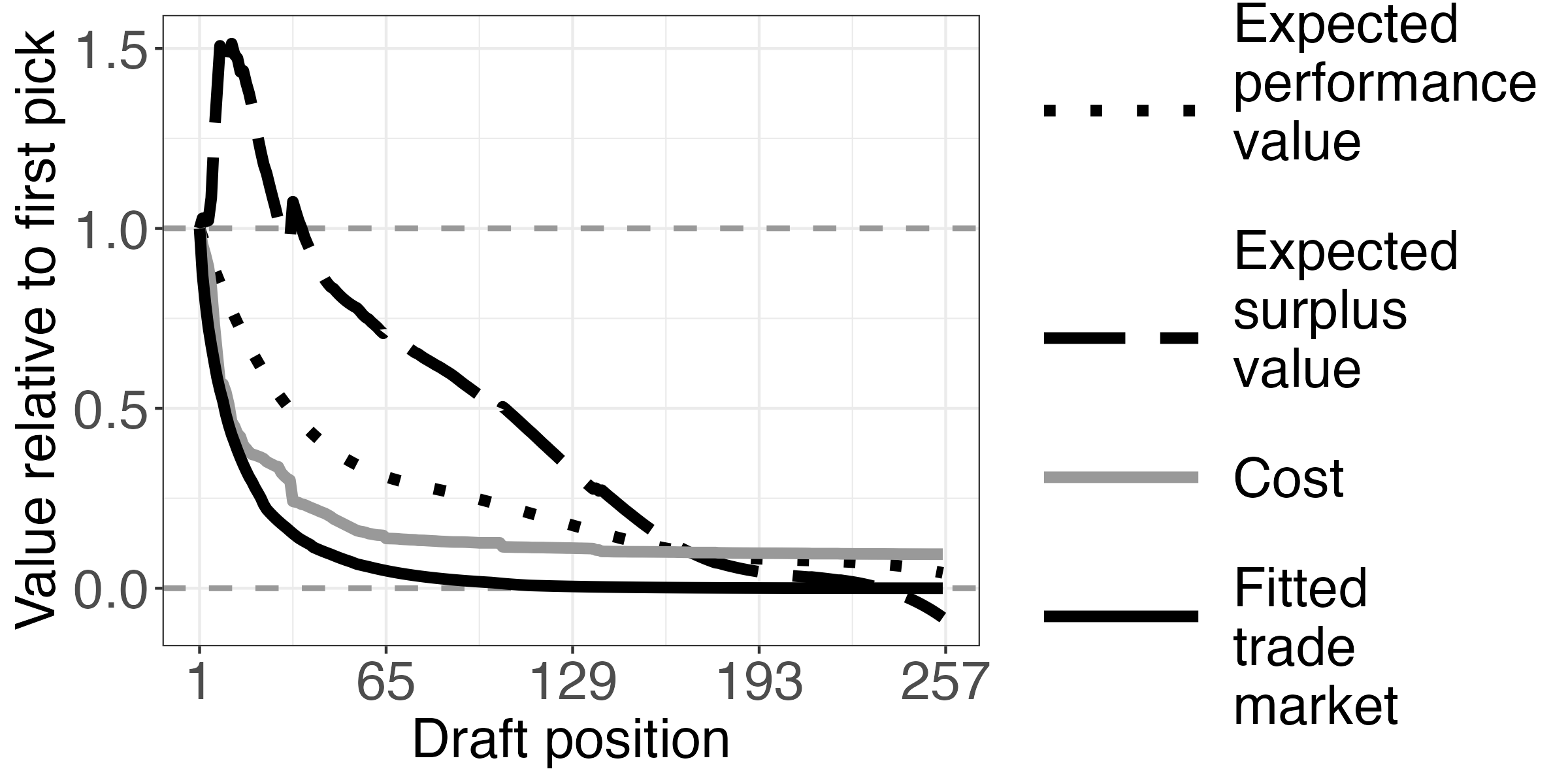}
    \caption{
    Each curve is a valuation of draft position $x$ relative to the first pick.
    The dotted black line is the expected performance value curve, proportional to $x \mapsto \bE[Y|x]$.
    The solid gray line is the cost curve, proportional to $x \mapsto \cost(x)$.
    The long-dashed black line is the expected surplus value curve, proportional to $x \mapsto \bE[Y-\cost(x)|x]$.
    The solid black line is the draft position value curve implied by the trade market.
    }
    \label{fig:plot_Massey_Thaler_replication_1}
\end{figure}

We fit these traditional draft position value curves using our outcome variable $Y$ (second contract annualized cap percentage), which we visualize in Figure~\ref{fig:plot_Massey_Thaler_replication_1}.
We use spline regression to fit the expected performance value curve from a dataset of all draft picks from 2013 to 2023 \citep{nflreadr}.
We calculate the cost curve as in \citet{BaldwinPosCurves}: beginning with the actual 2023 salary cap (about $\$225$ million) and assuming a cap growth rate of 7\% over the next three seasons (2024, 2025, and 2026), we convert the total dollar amount of the rookie contract into a percentage of the cap. 
Then, for each draft position we average the compensation values across the four seasons.
We calculate the expected surplus value curve as the difference between the expected performance value curve and cost curve.
Finally, we normalize each of these curves so that value is relative to the first pick.
Because of this normalization, note that the expected surplus value curve in Figure~\ref{fig:plot_Massey_Thaler_replication_1} does not visually look like the expected performance value curve minus the cost curve.
In Figure~\ref{fig:plot_Massey_Thaler_replication_1} we also visualize a draft position value curve implied by the trade market.
Devised in \citet{MasseyThaler2013}, the fitted trade market curve is a singular draft position value curve that best fits the dataset of all observed trades.
We derive this curve in detail in Appendix~\ref{app:trade_market_value}.

%%%%%%%%%%%%%%%%%%
%%%%%%%%%%%%%%%%%%

Expected performance value is less steep than cost, indicating that players are underpaid on average in their first contracts.
This produces positive expected surplus value across most of the draft (the long-dashed black line is mostly positive).
Expected surplus value is larger than one throughout the first round, peaking in the middle of the first round.
This surplus spike is what \citet{MasseyThaler2013} assert is the ``loser's curse'': top draft picks, belonging to last year's worst teams, produce less surplus value on average than draft picks later in the first round that belong to better teams.

The trade market curve (solid black line) is much steeper than the expected performance value curve (dotted black line) and the expected surplus value curve (long-dashed black line).
This implies that teams aren’t valuing draft positions by expected performance value or expected surplus value. 
Alternatively, if general managers are in fact adopting either of these as the utility function, they are trading suboptimally: they don't trade down often enough, and they often overpay when they trade up. 
According to expected surplus value, even teams at the very top of the draft should trade down, particularly towards the middle of the first round.
This result is counterintuitive because very top picks are so highly regarded by general managers and football fans.

%%%%%%%%%%%%%%%%%%%%%%%%%%%%%%%%%%%%%%%%%%%%%%%%%%%%%%%%%%%%%%%%%%%%%%%%%%%%%%%%%%%%%%%%%
\section{Research hypothesis}\label{sec:research_hypothesis}

\citet{MasseyThaler2013} developed the traditional analytical approach to constructing draft position value curves. 
Their study produced two key findings. 
First, they claimed to have found a loser's curse––last years worst teams are ``cursed'' with the ostensible gifts of the top picks, which cost a lot (in compensation) without delivering enough on-field value to compensate.
Second, they asserted that the draft pick trade market is highly inefficient––if one adopts the expected surplus value utility function, teams significantly overvalue top draft picks, consistently overpaying to trade up in the draft without trading down often enough.

Despite the widespread dissemination of these findings, general managers’ draft behavior has remained largely unchanged. 
% When applied to recent data (2013––2023), the loser's curse persists and the fitted trade market curve remains similar (see Section~\ref{sec:trad_value_chart} and \citeauthor[][forthcoming]{avery}). 
When applied to recent data (2013––2023), the loser's curse persists and the fitted trade market curve remains similar (see Section~\ref{sec:trad_value_chart}). 
If general managers were optimizing for expected performance value or expected surplus value, their continued reluctance to trade down and their willingness to pay a premium for higher picks would suggest systematic misjudgment. One possible explanation is that general managers have simply failed to learn from these analyses.
An alternative explanation is that general managers are instead operating under a fundamentally different utility function. If maximizing expected surplus value does not align with their actual objectives, then their observed behavior may be rational, even if it appears suboptimal under Massey and Thaler's framework. 
This paper challenges their core assumption, that general managers should adopt expected surplus value as their utility function in valuing draft picks.
This assumption is neither explicitly justified in their analysis nor necessarily optimal in the context of NFL decision making.

A general manager’s primary objective is typically considered to be constructing a Super Bowl-winning roster. 
Achieving this goal requires assembling a team capable of reaching and surpassing an elite performance threshold. Mathematically, thinking of team performance as a random variable, winning the Super Bowl is analagous to exceeding a high threshold of team performance relative to other teams. In such a setting, maximizing expected surplus value alone may not be optimal. A strategy that emphasizes high-variance outcomes—seeking players with the potential for extreme right-tail performance—may be preferable, even if it sacrifices consistency.

\begin{figure}[hbt!]
    \centering{}
    \includegraphics[width=\textwidth]{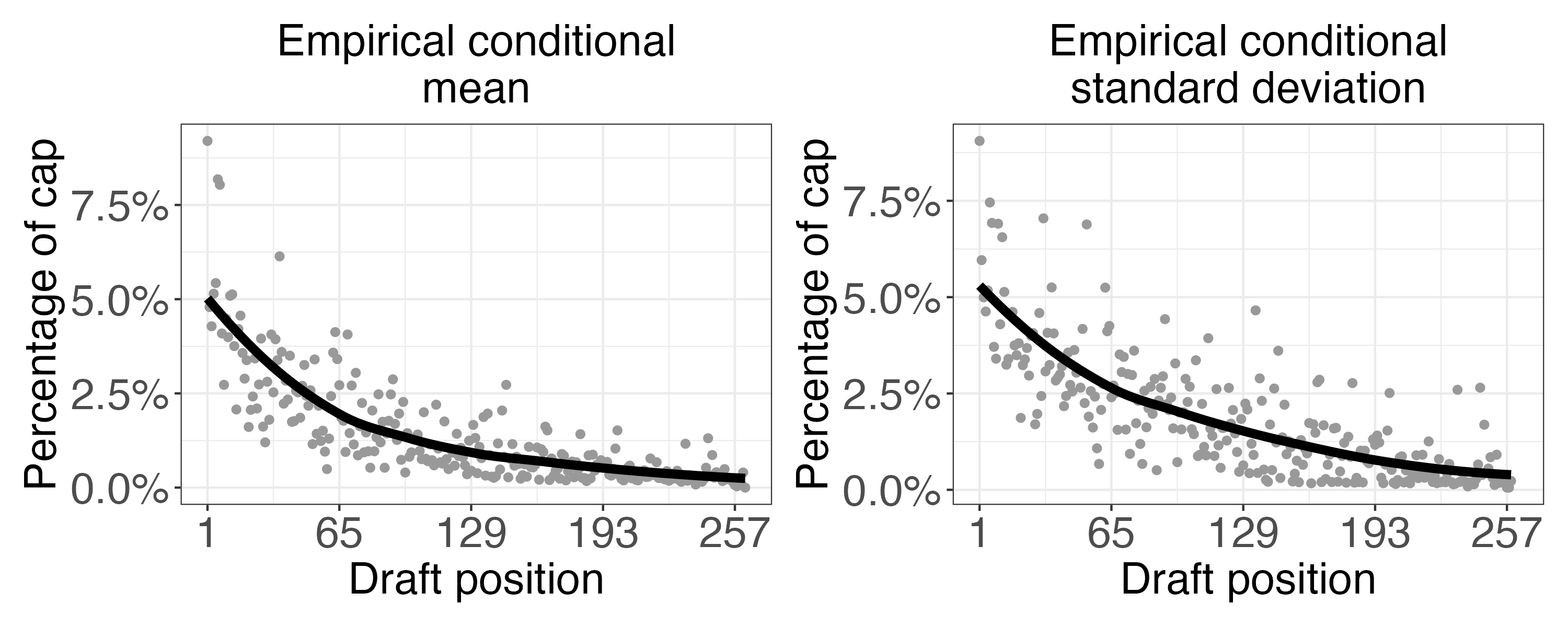}
    \caption{
        The gray dots denote the empirical mean (left) and standard deviation (right) of performance $Y$ given draft position $x$.
        The black lines are smoothed curves $x \mapsto \bE[Y|x]$ (left) and $x \mapsto \sd(Y|x)$ (right).
    }
    \label{fig:plot_empMeanSd}
\end{figure}

This framework has direct implications for evaluating draft picks. 
If elite players––those in the far right tail of the performance distribution––are disproportionately valuable in constructing a championship team, then upweighting variance in draft selections is rational. 
Figure~\ref{fig:plot_empMeanSd} shows that both the expected performance value and variance of performance decreases as draft position increases. 
Critically, the decline in variance suggests that higher draft picks offer significantly greater potential for elite outcomes, reinforcing the rationale for teams to value them more than suggested by expected surplus value models.

%%%%%%%%%%%%%%%%%%
%%%%%%%%%%%%%%%%%%

Thus, rather than viewing general managers’ draft strategies as systematically flawed, we propose that their behavior aligns with a decision rule that accounts for variance and right-tail outcomes. 
Our hypothesis is that, in trading draft picks, teams prioritize something closer to the probability of drafting transformational players over maximizing expected surplus value. 
These alternative valuation functions lead to steeper draft position value curves that are more consistent with observed trade behavior. 

%%%%%%%%%%%%%%%%%%%%%%%%%%%%%%%%%%%%%%%%%%%%%%%%%%%%%%%%%%%%%%%%%%%%
%%%%%%%%%%%%%%%%%%%%%%%%%%%%%%%%%%%%%%%%%%%%%%%%%%%%%%%%%%%%%%%%%%%%
%%%%%%%%%%%%%%%%%%%%%%%%%%%%%%%%%%%%%%%%%%%%%%%%%%%%%%%%%%%%%%%%%%%%
\section{Conditional density estimation}\label{app:estimate_con_density}

By modeling the distribution of performance as a function of relevant covariates, we can construct a broad class of draft position value curves that extend beyond expected performance value or expected surplus value.
Hence, in this section, we estimate the conditional density of performance given draft position (Section~\ref{app:estimate_con_density_nopos}) and given both draft position and player position (Section~\ref{app:estimate_con_density_pos}).

%%%%%%%%%%%%%%%%%%%%%%%%%%%%%%%%%%%%%%%%%%%%%%%%%%%%%%%%%%%%%%%%%%%%
\subsection{Position-agnostic conditional density}\label{app:estimate_con_density_nopos}

\begin{figure}[hbt!]
    \centering{}
    \includegraphics[width=\textwidth]{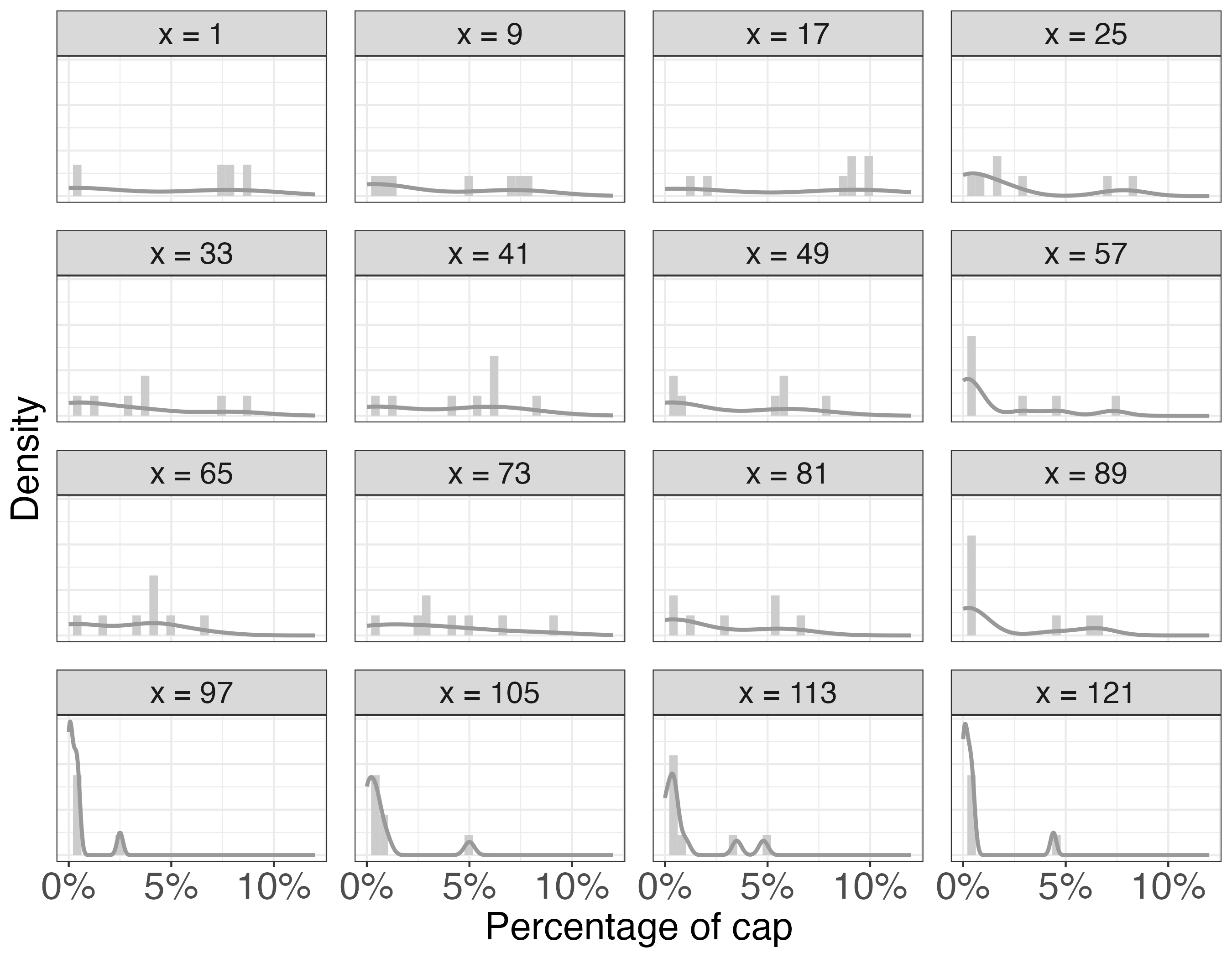}
    \caption{
        Empirical distribution (histogram and smoothed density) of performance $Y$ ($x$-axis) given draft position $x$ (facet).
    }
    \label{fig:plot_density_empirical_EDA}
\end{figure}

In Figure~\ref{fig:plot_density_empirical_EDA} we visualize the empirical conditional density of performance $Y$ given draft position $x$.
The empirical densities are extremely wobbly and noisy due to limited data for each individual draft position.
But there is a general trend: we see a spike near zero that grows upwards as draft position increases and a right tail that shifts leftwards and diminishes as draft position increases.
This motivates our ``Bayesian spike plus Beta regression'' model of $\bP(Y|x)$.\footnote{
    We initially tried a lone Beta regression model without the spike near zero, which less successfully fit the right tail.
}

%%%%%%%%%%%%%%%%%%
%%%%%%%%%%%%%%%%%%

We model the spike near zero, the \textit{bust spike}, using logistic regression.
We let a \textit{bust} be a player whose performance outcome $Y$ is smaller than some cutoff $\ybust$.
We model the bust probability $\bp(x) := \bP(Y \leq \ybust | x)$ by
\begin{equation}
    1\{Y\leq\ybust\} | x \sim \Bernoulli(\bp(x)),
\label{eqn:bust_prob_oerall_bernoulli}
\end{equation}
where 
\begin{equation}
    \bp(x) = \logistic(\alpha_0 + \alpha_1\cdot x).
\label{eqn:bust_prob_overall}
\end{equation}
Next, we model the right tail using Beta regression.
We view the Beta distribution as a natural way to model $Y$ because it lies in $[0,1]$ and aligns well with the empirical densities.
We model
\begin{equation}
    Y | x, Y>\ybust \sim \Beta(\mu(x), \phi(x)).
\label{eqn:beta_reg_overall}
\end{equation}
Here, we use the mean-precision parameterization of the Beta distribution.
The mean $\mu$ and precision $\phi$ are related to the traditional Beta distribution parameters $\shapeone$ and $\shapetwo$ by $\shapeone = \mu\cdot\phi$ and $\shapetwo = (1-\mu)\cdot\phi$.
We let the mean and precision parameters vary as draft position $x$ varies via $\mu(x)$ and $\phi(x)$ to capture that different draft positions $x$ yield different outcome distributions.
The conditional mean is $\bE[Y|x,Y>\ybust] = \mu(x)$ and the conditional variance is $\Var[Y|x,Y>\ybust] = \mu(x)\cdot(1-\mu(x))/(1+\phi(x))$.
So, the precision is inversely proportional to the variance.
We use spline regression to model the conditional mean,
\begin{equation}
    \mu(x) = \logistic(\widetilde{x}^\top \beta),
\label{eqn:beta_reg_overall_mu}
\end{equation}
where $\widetilde{x}$ is a B-spline basis.\footnote{
    The $\R$ code we used to generate the spline basis is \texttt{splines::bs($x$, degree=3, df=4, intercept=T)}.
}
We model the conditional precision by
\begin{equation}
    \phi(x) = \exp(\gamma_0 + \gamma_1\cdot x).
\label{eqn:beta_reg_overall_phi}
\end{equation}

% We use a bust spike in our model because without it

\begin{figure}[hbt!]
    \centering{}
    \includegraphics[width=\textwidth]{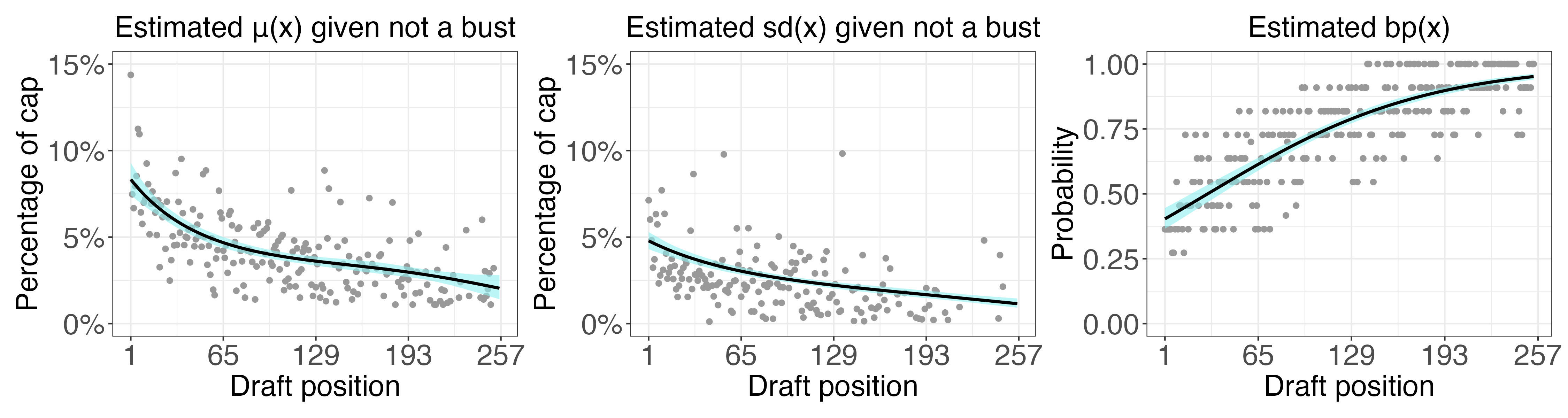}
    \caption{
        Left: our estimate of the conditional mean $x \mapsto \mu(x) = \bE[Y|x,Y>\ybust]$ given not a bust. 
        Middle: our estimate of the conditional standard deviation $x \mapsto \sd(x) = \sd(Y|x,Y>\ybust)$ given not a bust.
        Right: our estimate of bust probability $x \mapsto \bp(x) = \bP(Y\leq\ybust|x)$.
        The solid black lines are the posterior means, the cyan shaded regions are the $95\%$ credible intervals, and the gray dots are the empirical mean, empirical standard deviation, and empirical proportion values, respectively, for each draft position $x$.
    }
    \label{fig:plot_empWithCondLines}
\end{figure}

We use $\ybust = 0.5\%$ and model the log-odds of the mean $\mu(x)$ using a spline, the log of the precision $\phi(x)$ with a line, and the log-odds of the bust probability $\bp(x)$ with a line because these parametrizations fit the data well––for instance, Figure~\ref{fig:plot_empWithCondLines} shows that the fitted curves closely track the trends of the empirical curves.
Other values of $\ybust$ produced poorer fits (e.g., the fitted curves in Figure~\ref{fig:plot_empWithCondLines} tracked the empirical curves less closely).
Modeling the precision with a spline led to overfitting, and using a spline for the bust probability produced negligible improvements over the linear model.

To capture uncertainty in the parameters, we use a Bayesian model, placing priors on the parameters.
We put a weakly informative Normal prior with mean zero and standard deviation ten on each of the $\alpha$, $\beta$, and $\gamma$ parameters.
Finally, we assume that $Y$ is uniformly distributed within the bust spike in a way that doesn't vary with $x$: $Y | x, Y\leq\ybust \sim \Unif[0,\ybust]$.\footnote{
    We assess the reasonableness of this assumption later in this section.
}

Because the posterior distribution of $(\alpha, \beta, \gamma)$ is not analytically tractable, we use Markov Chain Monte Carlo (MCMC) to draw approximate samples from the posterior distribution.
We implement our sampler in \textbf{Stan} \citep{Stan} and perform our MCMC simulation using the \textbf{rstan} package \citep{rstan}.
To obtain our posterior samples, we run one MCMC chain for 2,500 iterations.
After discarding the first 1,250 iterations of each chain as ``burn-in'', the Gelman-Rubin $\hat{R}$ statistic for each parameter is less than $1.1$, suggesting convergence \citep{GelmanRubin1992}.

\begin{figure}[hbt!]
    \centering{}
    \includegraphics[width=\textwidth]{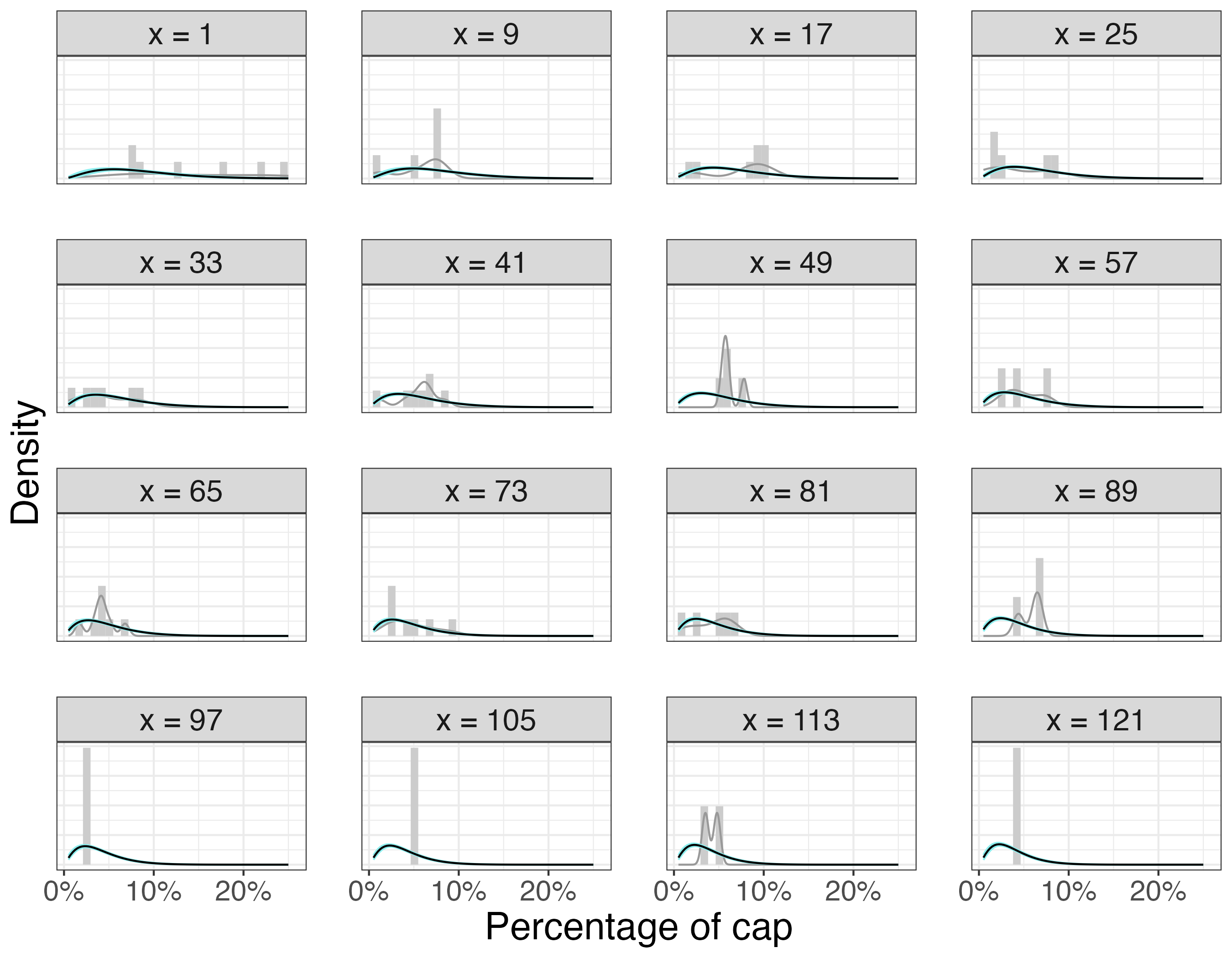}
    \caption{
        The distribution of performance $Y$ ($x$-axis) given draft position $x$ (facet). The solid black lines are the fitted conditional densities $\bP(Y|x,Y>\ybust)$ (the posterior mean curves) and the cyan shaded regions are the $95\%$ credible intervals, which overlay the empirical distributions (histograms and smoothed densities) in gray.
    }
    \label{fig:plot_density_empirical_fitted}
\end{figure}

\begin{figure}[hbt!]
    \centering{}
    \includegraphics[width=\textwidth]{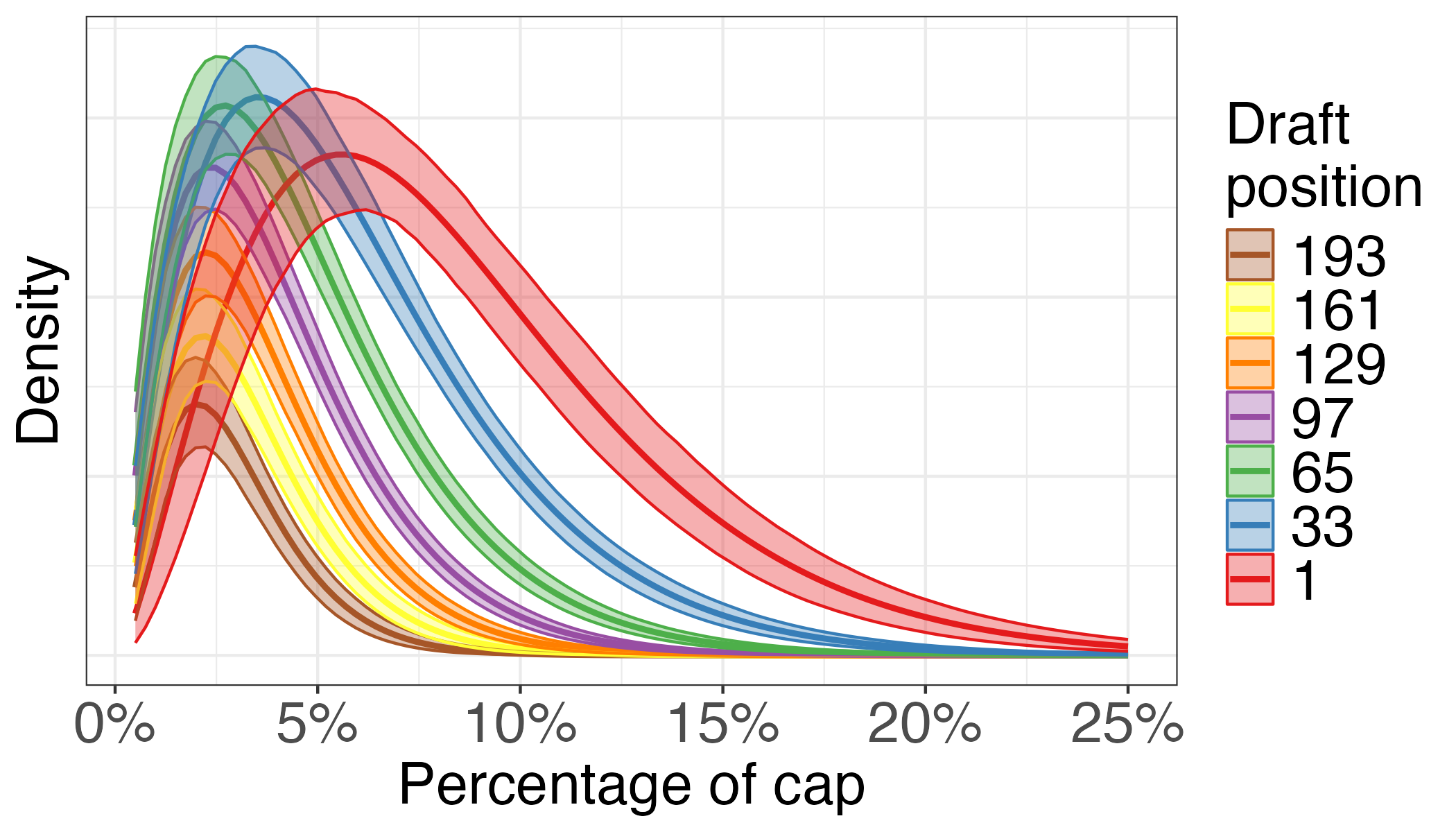}
    \caption{
    For values $Y$ in the right tail $Y > \ybust$, the estimated conditional density $\bP(Y|x) = \bP(Y|x,Y>\ybust) \cdot (1-\bp(x))$ for various values of draft position $x$ (color).
    The lines are posterior mean density curves and the shaded regions are 95\% credible intervals of those curves.
    }
    \label{fig:plot_post_density_rdsall}
\end{figure}

\begin{figure}[hbt!]
    \centering{}
    \subfloat[\centering ]{
        {\includegraphics[width=0.5\textwidth]{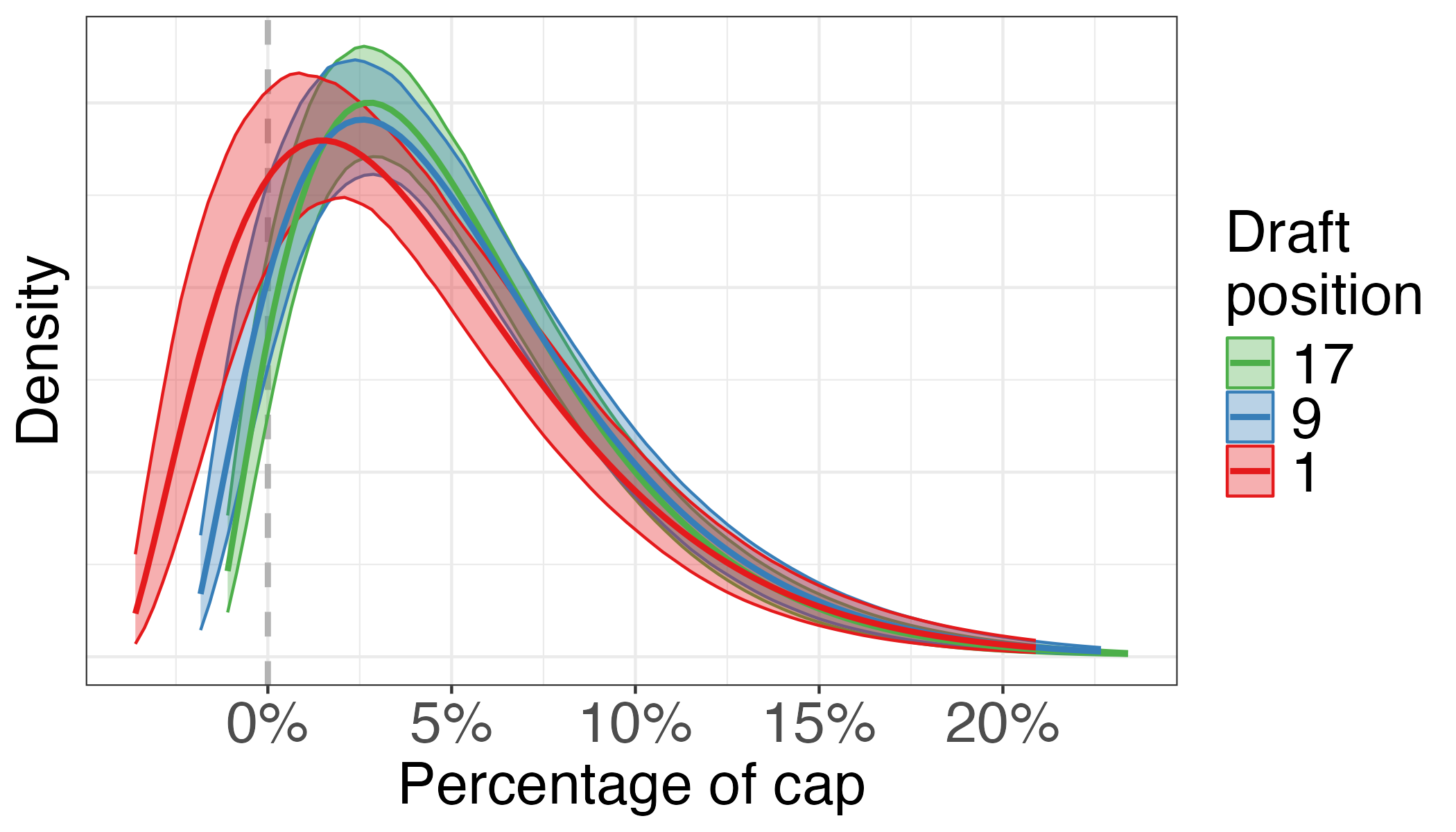}}
        \label{fig:plot_post_surplus_density_full_rd1}
    }
    % \qquad
    \subfloat[\centering ]{
        {\includegraphics[width=0.5\textwidth]{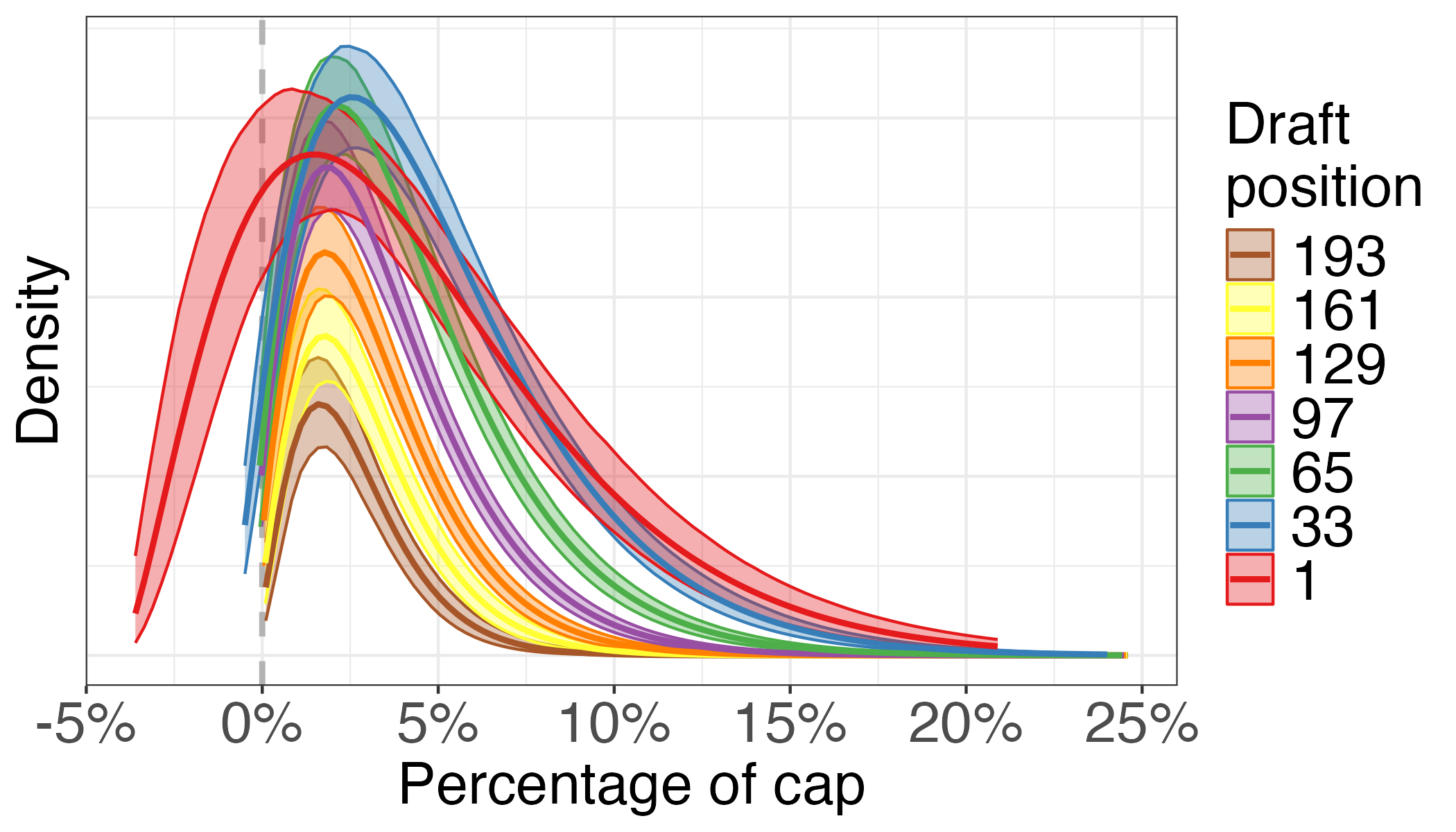}}
        \label{fig:plot_post_surplus_density_full_rdsall}
    }
    \caption{
    For values $Y$ in the right tail $Y > \ybust$, the estimated conditional surplus density $\bP(S|x) = \bP(Y-\cost(x)|x,Y> \ybust) \cdot (1-\bp(x))$ for various values of draft position $x$ (color) in the first round (a) and throughout the draft (b).
    The lines are posterior mean density curves and the shaded regions are 95\% credible intervals of those curves.
    }
    \label{fig:plot_post_surplus_density_full}
\end{figure}

In Figure~\ref{fig:plot_empWithCondLines} we see that our estimates of the conditional mean $\mu(x)$, conditional standard deviation $\sd(x)$, and bust probability $\bp(x)$ fit the data well. 
In Figure~\ref{fig:plot_density_empirical_fitted} we see that our estimates of the right tail $\bP(Y|x,Y>\ybust)$ fit the data well enough.
As expected, the right tail shifts leftwards as draft position increases.
In Figure~\ref{fig:plot_post_density_rdsall} we visualize the conditional density $\bP(Y|x)$ for values $Y$ in the right tail $Y > \ybust$.
By Bayes rule, this is equal to $\bP(Y|x,Y>\ybust) \cdot (1-\bp(x))$.
The right tail, measuring the probability of drafting an elite player, shifts leftwards and decreases in size as draft position increases.
In Figure~\ref{fig:plot_post_surplus_density_full} we visualize the conditional density of surplus $S = Y - \cost(x)$.
The high cost of first round picks shifts those curves so much to the left that there is substantial probability mass on negative (below-replacement) outcomes.

%%%%%%%%%%%%%%%%%%
%%%%%%%%%%%%%%%%%%

%%%%%%%%%%%%%%%%%%
%%%%%%%%%%%%%%%%%%

%%%%%%%%%%%%%%%%%%
%%%%%%%%%%%%%%%%%%

%%%%%%%%%%%%%%%%%%
%%%%%%%%%%%%%%%%%%

\begin{figure}[hbt!]
    \centering{}
    \includegraphics[width=\textwidth]{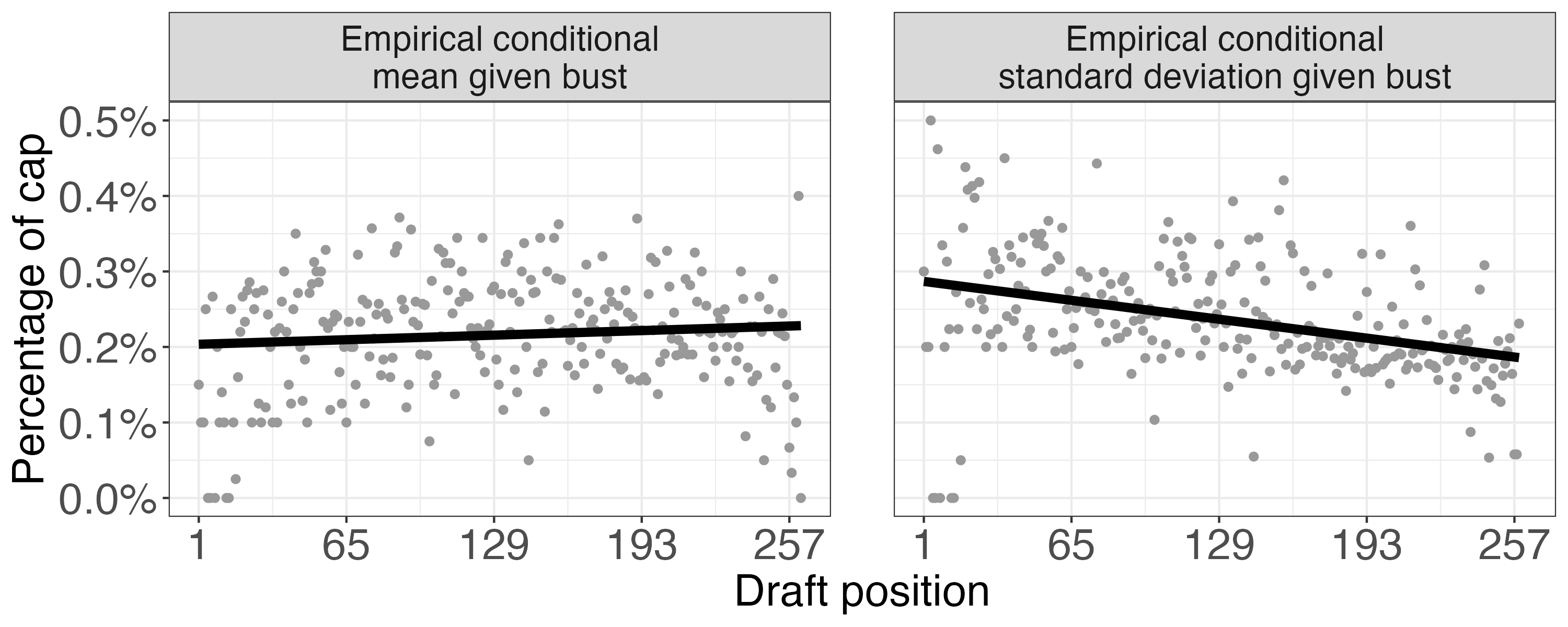}
    \caption{
        The gray dots display the empirical mean (left) and standard deviation (right) of performance $Y$ given draft position $x$ conditional on being a bust $Y \leq \ybust$. The black lines are smoothed lines.
    }
    \label{fig:plot_empBustSpikeDist}
\end{figure}

Finally, in Figure~\ref{fig:plot_empBustSpikeDist} we assess the assumption that the bust spike conditional distribution of $Y|x,Y\leq\ybust$ is uniform in a way that doesn't vary with $x$.
The empirical conditional mean stays essentially constant over $x$ and the empirical conditional variance slightly decreases linearly in $x$.
This admittedly renders the uniform assumption imperfect.
Nonetheless, we ultimately feel it is good enough.
The assumption is benign because when $Y \leq \ybust$, $Y$ is essentially 0 and so negligbly impacts the right tail probability calculations that form the draft trade value curves that we detail in this paper.

%%%%%%%%%%%%%%%%%%
%%%%%%%%%%%%%%%%%%

%%%%%%%%%%%%%%%%%%%%%%%%%%%%%%%%%%%%%%%%%%%%%%%%%%%%%%%%%%%%%%%%%%%%
%%%%%%%%%%%%%%%%%%%%%%%%%%%%%%%%%%%%%%%%%%%%%%%%%%%%%%%%%%%%%%%%%%%%
\subsection{Position-specific conditional density}\label{app:estimate_con_density_pos}

% Now, we model the conditional density of performance outcome $Y$ given draft position $x$ and position $\pos$.
% We extend
Now, we extend the ``Bayesian spike plus Beta regression'' model from the previous section to adjust for player position, henceforth referred to as just ``position'' and denoted $\pos$.
We use 11 position groups: quarterback (QB), wide receiver (WR), running back / fullback (RB/FB), tight end (TE), offensive tackle (OT), interior offensive lineman (IOL), interior defensive lineman (DL), edge (ED), linebacker (LB), cornerback (CB), and safety (S).
We filter out all draft picks from our dataset belonging to other positions (e.g., punter, kicker, etc.).

It is imperative to use a Bayesian model to adjust for position, rather than fitting separate curves for each position, due to limited data.
Across each of the positions, the median sample size in our dataset is 269 players, with a maximum of 350 and a minimum of 127.
Spread across 256 total draft pick positions, we are left with an average of about one observation per combination of position and draft position.
A Bayesian hierarchical model is an excellent choice here because it allows us to share information across observations via shrinkage and to capture uncertainty in position curves.

We devise a Bayesian hierarchical model that includes position-specific parameters for each of the $\alpha$, $\beta$, and $\gamma$ parameters from the previous model. 
% position-agnostic model from Appendix~\ref{app:estimate_con_density_nopos}.
We model the right tail conditional density by
\begin{equation}
    Y | x, \pos, Y>\ybust \sim \Beta(\mu(x,\pos), \phi(x,\pos)),
\label{eqn:beta_reg_pos}
\end{equation}
where
\begin{equation}
    \mu(x,\pos) = \logistic(\widetilde{x}^\top \beta_\pos)
\label{eqn:beta_reg_pos_mu}
\end{equation}
and
\begin{equation}
    \phi(x,\pos) = \exp(\gamma_{0,\pos} + \gamma_{1,\pos}\cdot x).
\label{eqn:beta_reg_pos_phi}
\end{equation}
As before, $\widetilde{x}$ is a B-spline basis on the $x$ variable.
We model bust probability by
\begin{equation}
    \bp(x,\pos) := \bP( Y\leq\ybust | x, \pos) = \logistic(\alpha_{0,\pos} + \alpha_{1,\pos}\cdot x).
\label{eqn:bust_prob_pos}
\end{equation}
We shrink the position-specific coefficients $\alpha_{\ast,\pos}$, $\beta_{\ast,\pos}$, and $\gamma_{\ast,\pos}$ towards overall mean parameters $\alpha_\ast$, $\beta_\ast$, and $\gamma_\ast$.
For example, we model the position-specific coefficients $\alpha_{0,\pos}$ as a draw from a Normally distributed prior with mean $\alpha_0$ and standard deviation $\tau_{\alpha_0}$.
We put a weakly informative Normal prior with mean zero and standard deviation ten on the prior mean parameter $\alpha_0$.
We put a mild positive standard Normal prior on the prior standard deviation parameter $\tau_{\alpha_0}$.
The priors and hyperpriors are the same for the other parameters.

We again use \textbf{Stan} to draw approximate posterior samples of the parameters $(\alpha, \beta, \gamma)$, running one chain for 50,000 iterations.
% To obtain our posterior samples, we run one MCMC chain for 50,000 iterations.
After discarding the first 25,000 iterations of each chain as ``burn-in'', the Gelman-Rubin $\hat{R}$ statistic for each parameter is less than $1.1$, suggesting convergence \citep{GelmanRubin1992}.
% The Gelman-Rubin $\hat{R}$ statistic for each parameter is less than $1.1$, suggesting convergence \citep{GelmanRubin1992}.

\begin{figure}[p]
    \centering{}
    \subfloat[\centering ]{
        {\includegraphics[width=0.92\textwidth]{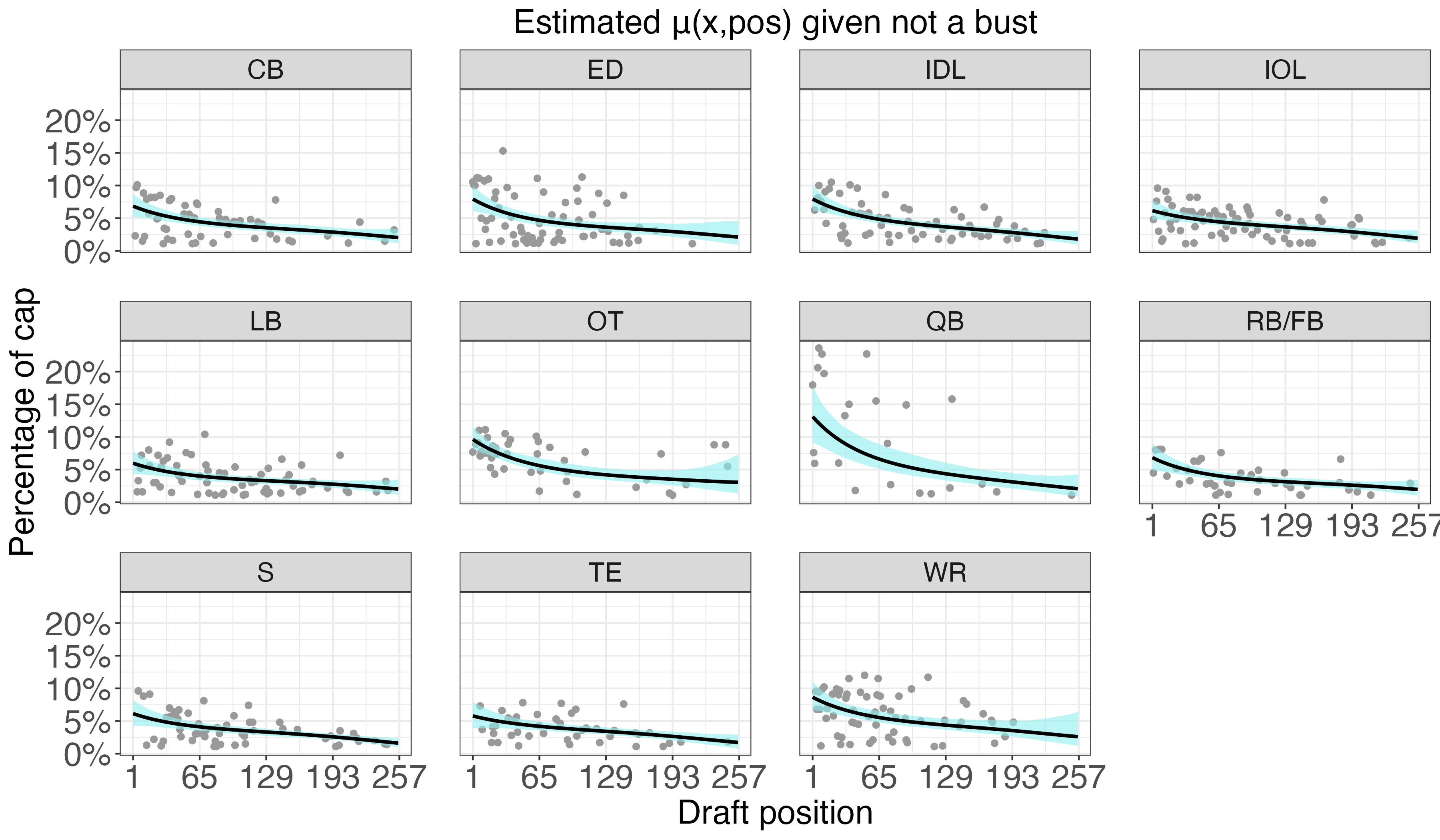}}
        \label{fig:plot_empWithCondMean_byPos}
    }
    \\
    \subfloat[\centering ]{
        {\includegraphics[width=0.92\textwidth]{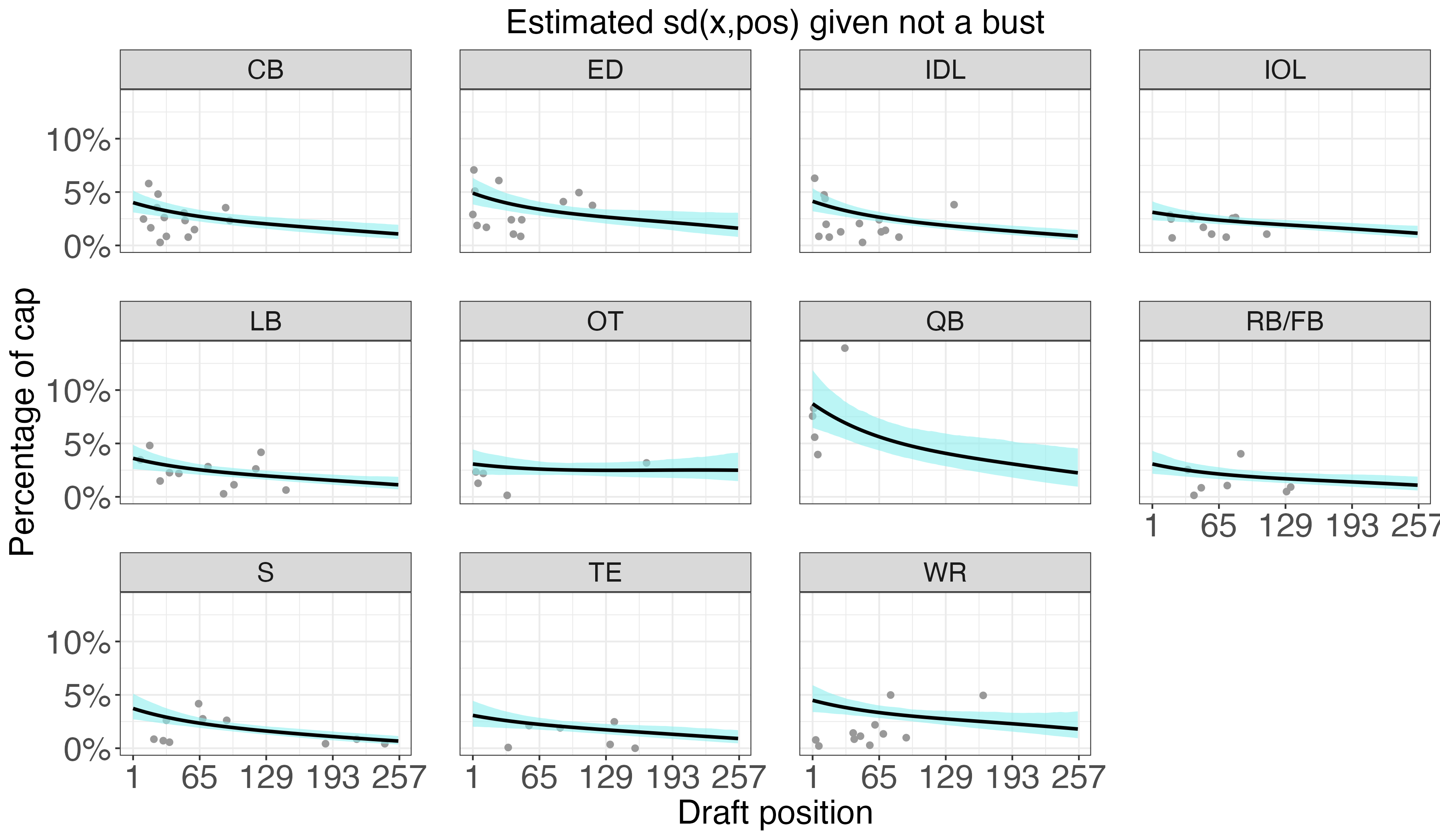}}
        \label{fig:plot_empWithCondSd_byPos}
    }
    \caption{
    Our estimates of the conditional mean $\mu(x,\pos) = \bE[Y|x,\pos,Y>\ybust]$ (a) and standard deviation $\sd(x,\pos) = \sd(Y|x,\pos,Y>\ybust)$ (b) given not a bust. 
    The solid black lines are the posterior means, the cyan shaded regions are the $95\%$ credible intervals, and the gray dots are the empirical base-rates.
    }
    \label{fig:plot_empWithCondLines_byPos}
\end{figure}

%%% FIGURES
In Figure~\ref{fig:plot_empWithCondMean_byPos} we see that our estimate of the conditional mean $\mu(x,\pos)$ fits the data well.
% The fitted posterior means match the data almost perfectly.
The empirical standard deviation values in Figure~\ref{fig:plot_empWithCondSd_byPos} we are much more sparse than the empirical conditional mean values in Figure~\ref{fig:plot_empWithCondMean_byPos}.
This is because, for each position $\pos$ and for many draft positions $x$, just zero or one player of position $\pos$ was drafted at $x$.
Thus, there is no empirical standard deviation at many combination of $x$ and $\pos$.
Nonetheless, based on the data we have observed, our estimate of the conditional standard deviation $\sd(x,\pos)$ fits the data reasonably well.

%%%%%%%%%%%%%%%%%%
%%%%%%%%%%%%%%%%%%

\begin{figure}[p]
    \centering
    \subfloat[\centering ]{
        {\includegraphics[width=\textwidth]{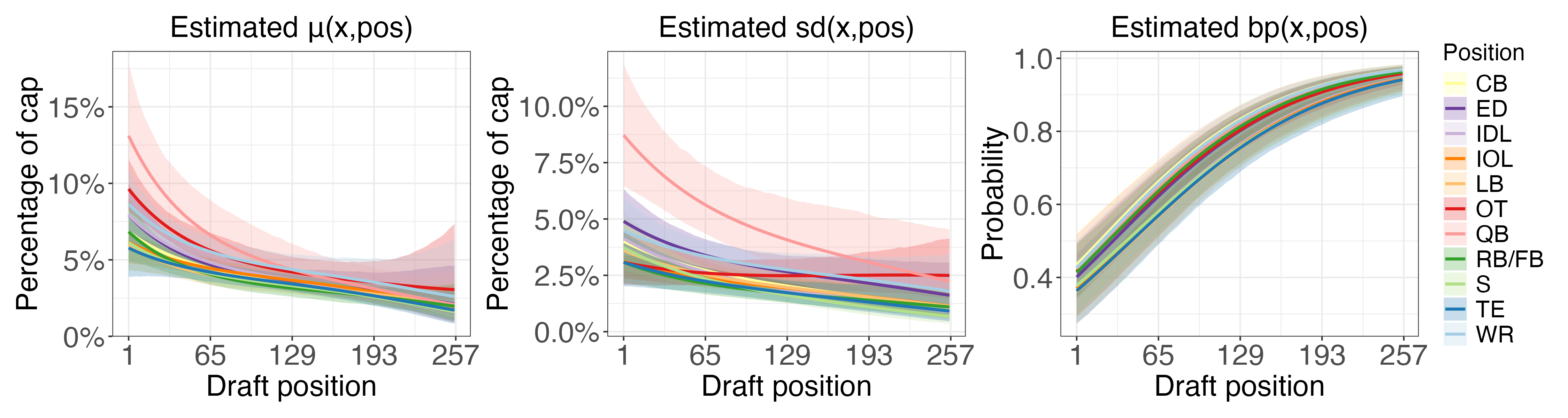}}
        \label{fig:plot_condLinesSE_byPos}
    }
    \\
    \subfloat[\centering ]{
        {\includegraphics[width=\textwidth]{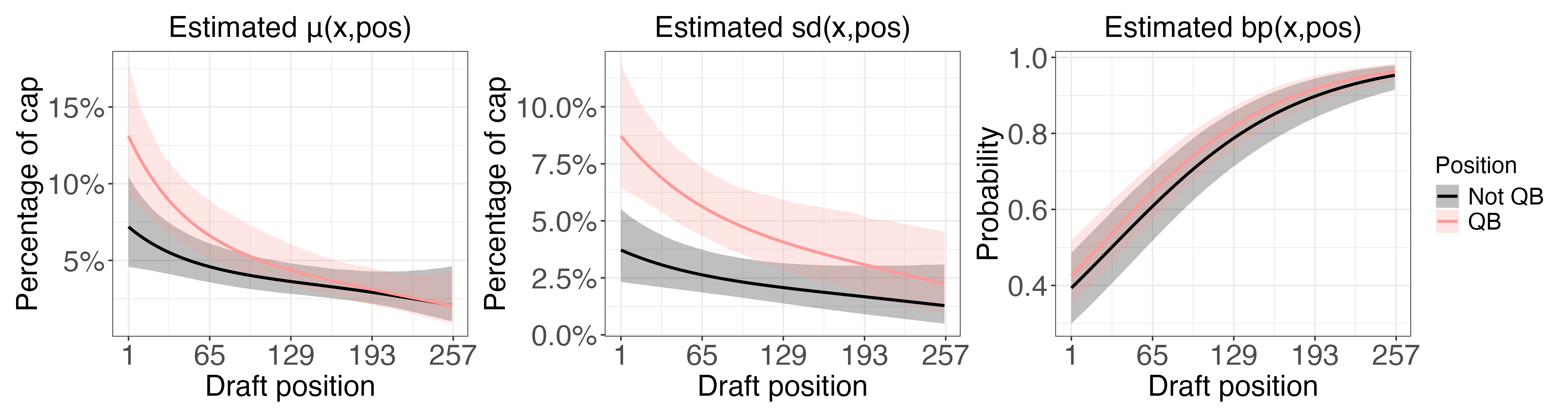}}
        \label{fig:plot_condLinesSE_byQB}
    }
    \caption{
    Figure (a): The posterior mean estimates (lines) and 95\% credible intervals (shaded regions) of $\mu(x,\pos)$ (left), $\sd(x,\pos)$ (middle), and $\bp(x,\pos)$ (right) as a function of draft position $x$ and position $\pos$ (color).
    Figure (b): These curves for $\pos \in \{\qb, \notqb\}$.
    }
    \label{fig:plot_ThreeCondLinesByPos}
    %%%%%%%%%%%%%%%%%%%%%%%%%%%%%%%%%%%%%%%%%%%%%%%%%%%%%%%%%%%%%%%%%%%%%%%%
    \vspace{2cm} % Adds space between the figures
    %%%%%%%%%%%%%%%%%%%%%%%%%%%%%%%%%%%%%%%%%%%%%%%%%%%%%%%%%%%%%%%%%%%%%%%%
    \includegraphics[width=\textwidth]{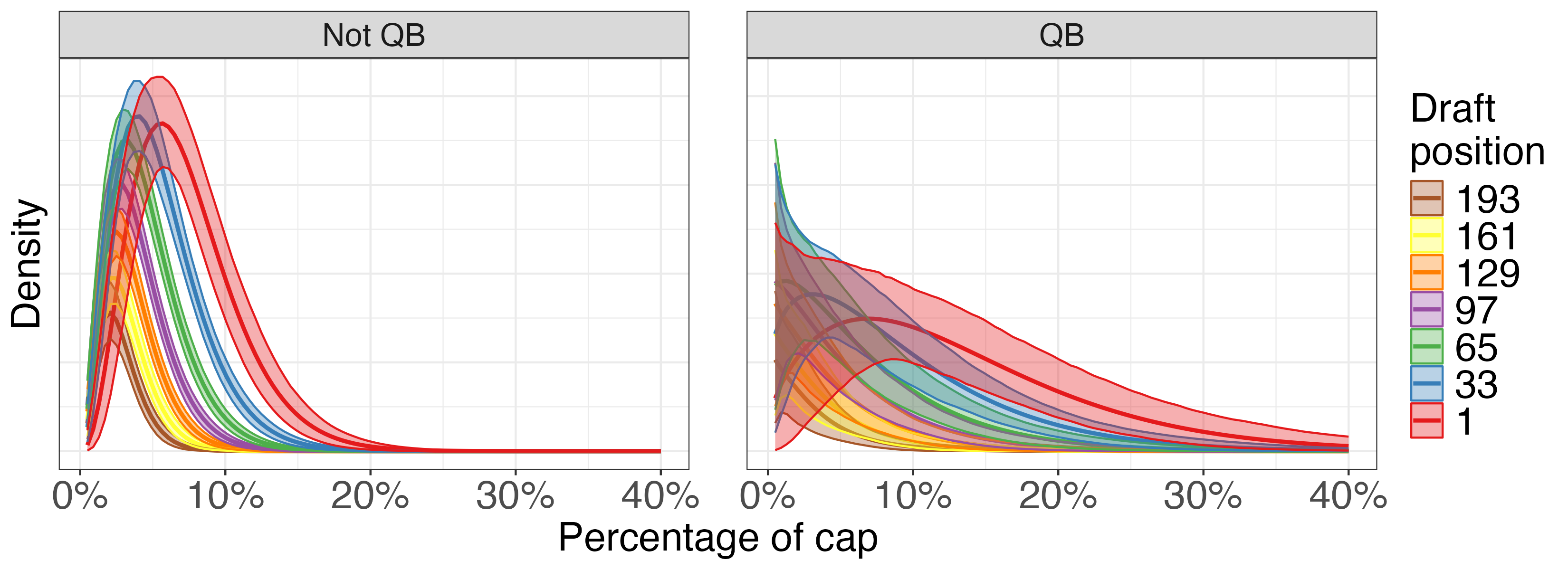}
    \caption{
        For values of $Y$ in the right tail $Y > \ybust$, the posterior mean estimates (lines) and 95\% credible intervals (shaded regions) of the conditional density $\bP(Y|x,\pos) = \bP(Y|x,\pos,Y>\ybust) \cdot (1-\bp(x,\pos))$, for $\pos \in \{\qb, \notqb\}$ (facet) and various values of draft position $x$ (color) throughout the draft.
    }
    \label{fig:plot_post_density_full_byQB}
\end{figure}

We see in Figure~\ref{fig:plot_empWithCondLines_byPos} that the quarterback mean and standard deviation curves stand out relative to every other position.
In Figure~\ref{fig:plot_ThreeCondLinesByPos}, we view this through a different lens.
In Figure~\ref{fig:plot_condLinesSE_byPos} we visualize our estimates of the conditional mean, standard deviation, and bust probability given position and draft position.
Though the posterior mean curves vary by position, the posterior credible intervals are wide.
The non-quarterback mean and standard deviation curves jumble into one big clump.
Hence, in Figure~\ref{fig:plot_condLinesSE_byQB} we average all of the non-quarterback curves into the black curves, $\mu(x,\notqb) = \frac{1}{\#\pos-1}\sum_{\pos' \neq \qb} \mu(x,\pos')$ and similarly for $\sd(x,\notqb)$ and $\bp(x,\notqb)$.
As expected, quarterbacks have a higher mean than other positions.
More interestingly, quarterbacks have a much higher variance than other positions.
This is what drives the fat far right tail of top-pick quarterback performance.

%%%%%%%%%%%%%%%%%%
%%%%%%%%%%%%%%%%%%

In Figure~\ref{fig:plot_post_density_full_byQB} we visualize the conditional density $\bP(Y|x,\pos)$ for $\pos \in \{\qb, \notqb\}$ (where the $\notqb$ density is the average of all the non-quarterback positional densities).
As the draft progresses, for both $\qb$'s and $\notqb$'s the right tail shifts leftwards and decreases in size as draft position increases.
Notably, the $\qb$ right tails are much fatter than the $\notqb$ tails.
This reflects the immense value of quarterbacks, and top-pick quarterbacks in particular.

%%%%%%%%%%%%%%%%%%%%%%%%%%%%%%%%%%%%%%%%%%%%%%%%%%%%%%%%%%%%%%%%%%%%%%%%%%%%%%%%%
%%%%%%%%%%%%%%%%%%%%%%%%%%%%%%%%%%%%%%%%%%%%%%%%%%%%%%%%%%%%%%%%%%%%%%%%%%%%%%%%%
%%%%%%%%%%%%%%%%%%%%%%%%%%%%%%%%%%%%%%%%%%%%%%%%%%%%%%%%%%%%%%%%%%%%%%%%%%%%%%%%%
\section{Results}\label{sec:results}

Given our estimated conditional density of performance given draft position and player position, in this section we construct draft position value curves from alternative utility functions.
We move beyond expected performance value and expected surplus value to utility functions that emphasize elite, right-tail outcomes.

%%%%%%%%%%%%%%%%%%%%%%%%%%%%%%%%%%%%%%%%%%%%%%%%%%%%%%%%%%%%%%%%%%%%%%%%%%%%%%%%%%%%%%%%%
\subsection{Right tail probability}\label{sec:right_tail_prob}

An intuitive approach to valuing a draft position in a way that prioritizes eliteness is by right-tail probability.
Formally, we consider the probability that the performance $Y$ of a player drafted at position $x$ exceeds a threshold $r$, $\bP(Y>r|x)$.
The threshold $r$
is determined relative to a percentage of the NFL salary cap.
For example, a general manager may be interested in the probability that a drafted player achieves at least the performance level of quarterback Derek Carr ($r=15.0\%$).
Since the expected value of an indicator function is its probability, this equivalently measures the number of elite players expected to arise from a given pick.

\begin{figure}[hbt!]
    \centering{}
    \includegraphics[width=\textwidth]{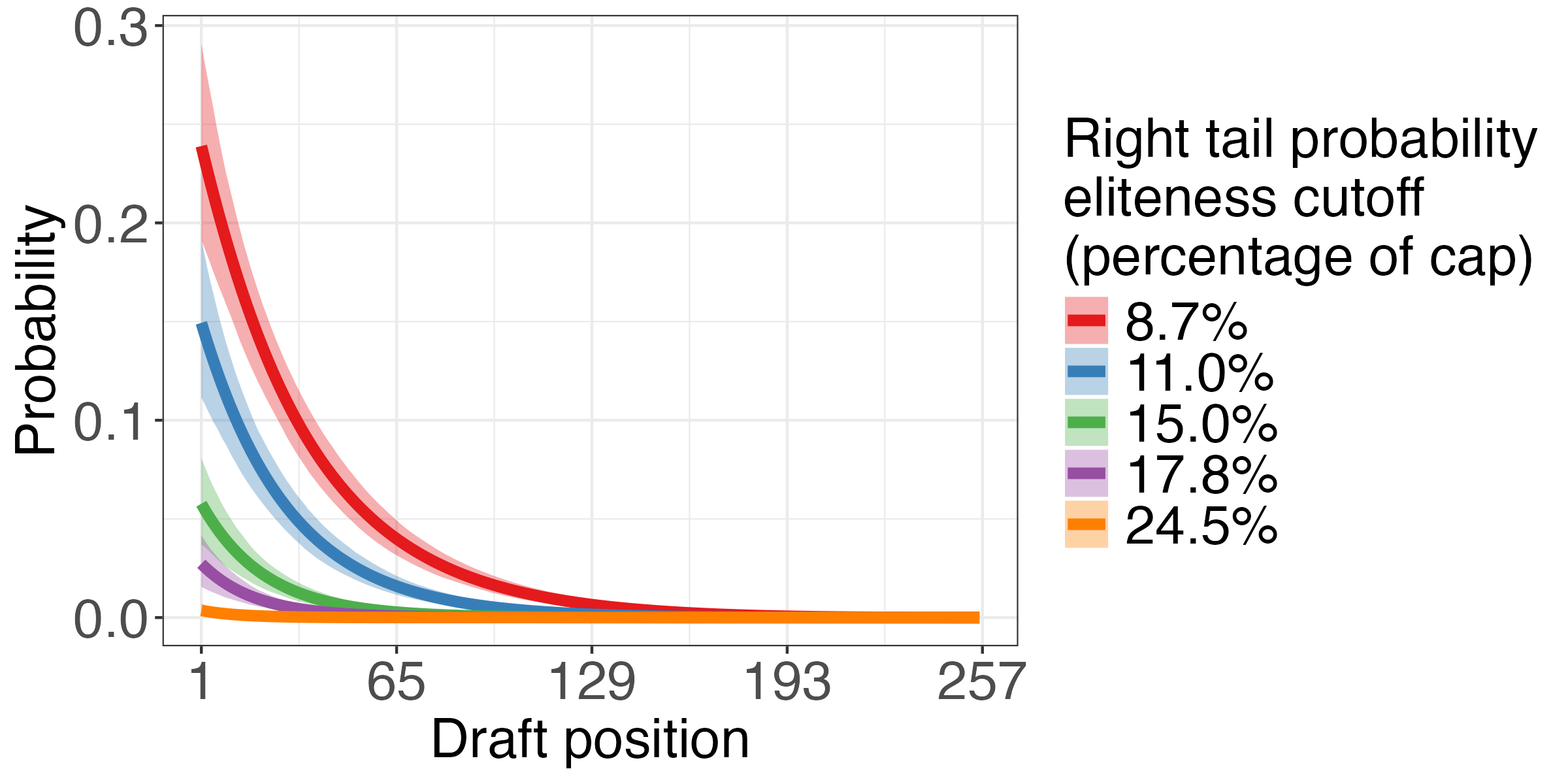}
    \caption{
        Estimated right tail probability $\bP(Y>r|x)$ ($y$-axis) as a function of draft position $x$ and eliteness cutoff $r$ (color).
        The lines are the posterior means and the shaded regions are the $95\%$ credible intervals.
    }
    \label{fig:plot_tail_probs_raw}
\end{figure}

We compute right tail probabilities from the estimated conditional density $\bP(Y|x)$. 
As both the mean and variance of $Y$ decrease convexly as the draft progresses (see Figure~\ref{fig:plot_empMeanSd})––because the densities of top picks feature fat right tails, which shift lefward and morph into a spike near zero as the draft progresses (see Figure~\ref{fig:plot_post_density_rdsall})––the right tail of the performance distribution exhibits a similar convex decay, which we visualize in Figure~\ref{fig:plot_tail_probs_raw}.
We consider five eliteness cutoffs $r$: $24.5\%$, $17.8\%$ $15.0\%$, $11.0\%$ and $8.7\%$.
Among quarterbacks in our dataset who secured multi-year second contracts as regular starters, Joe Burrow has the highest performance value ($r = 24.5\%$) and Derek Carr has the lowest ($r = 15.0\%$). 
Jared Goff lies near the midpoint of these two benchmarks ($r = 17.8\%$).
Among non-quarterbacks, offensive tackle Penei Sewell ranks at the $99^{th}$ percentile ($r = 11.0\%$), while cornerback Tre'Davious White lies at the $90^{th}$ percentile ($r = 8.7\%$).

%%%%%%%%%%%%%%%%%%
%%%%%%%%%%%%%%%%%%

%%%%%%%%%%%%%%%%%%
\begin{figure}[hbt!]
    \centering{}
    \includegraphics[width=\textwidth]{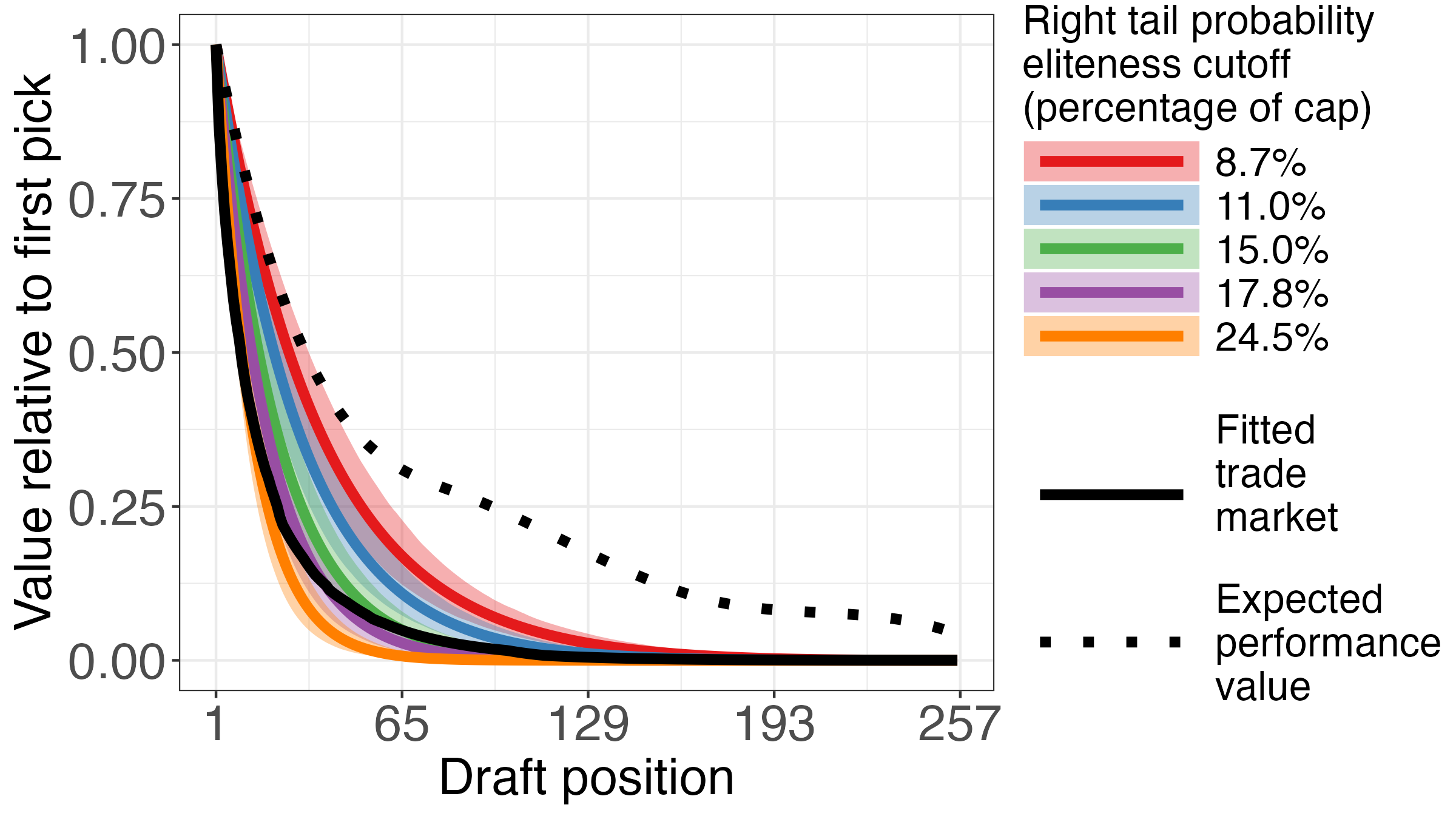}
    \caption{
        Value relative to the first pick ($y$-axis) as a function of draft position $x$. 
        The dotted black line is the expected performance value curve.
        The solid black line is the fitted trade market curve.
        The colored lines are proportional to right tail probability $x \mapsto \bP(Y>r|x)$, where $r$ is an eliteness cutoff (color).
        The lines are the posterior means and the shaded regions are the $95\%$ credible intervals.
    }
    \label{fig:plot_tail_probs}
\end{figure}
%%%%%%%%%%%%%%%%%%

In Figure~\ref{fig:plot_tail_probs} we visualize draft position value curves proportional to right tail probability, $x \mapsto \bP(Y > r|x)/\bP(Y > r|x=1)$.
We overlay the expected performance value curve and the fitted trade market curve from Figure~\ref{fig:plot_Massey_Thaler_replication_1}.
Notably, the right tail probability curves are steeper than the expected performance value curve. 
This suggests that if teams prioritize eliteness rather than expected performance value, earlier draft picks hold substantially greater value relative to later picks. 
Furthermore, as the eliteness threshold $r$ increases, right-tail draft position value curves progressively steepen, ultimately converging to or even surpassing the fitted trade market curve.

Further, we find that general managers trade draft picks as if they value a draft position by the probability it produces a player whose second contract exceeds $19.7\%$ of the salary cap.
Specifically, the mean absolute error between the fitted trade market curve and a right tail probability curve with eliteness threshold $r$ is minimized by $r=19.7\%$ (note that Deshaun Watson has $r=19.7\%$).
This is a high bar of eliteness: among players in our dataset who secured multi-year second contracts as regular starters, just $12.3\%$ of quarterbacks ($8/65$) and zero non-quarterbacks met this criterion.
Although we cannot directly observe general managers’ internal valuation models, the trade market aligns closely with what we would expect if teams were optimizing for elite player acquisition rather than expected performance value.

%%%%%%%%%%%%%%%%%%%%%%%%%%%%%%%%%%%%%%%%%%%%%%%%%%%%%%%%%%%%%%%%%%%%%%%%%%%%%%%%%%%%%%%%%
\subsection{Surplus value}\label{sec:surplus}

\citet{MasseyThaler2013} assess draft pick value through the lens of surplus value, the difference between a player’s performance during his first contract and the cost associated with his draft position (see Figure~\ref{fig:plot_Massey_Thaler_replication_1}).
Formally, denoting the first contract compensation at draft position $x$ by $\cost(x)$ and a drafted player's performance outcome by $Y$, surplus value is defined by $S = Y - \cost(x)$.
While \citet{MasseyThaler2013} modeled expected surplus value, $\bE[S|x]$, we consider the right tail of surplus value, $\bP(S>r|x)$.

\begin{figure}[hbt!]
    \centering{}
    \includegraphics[width=\textwidth]{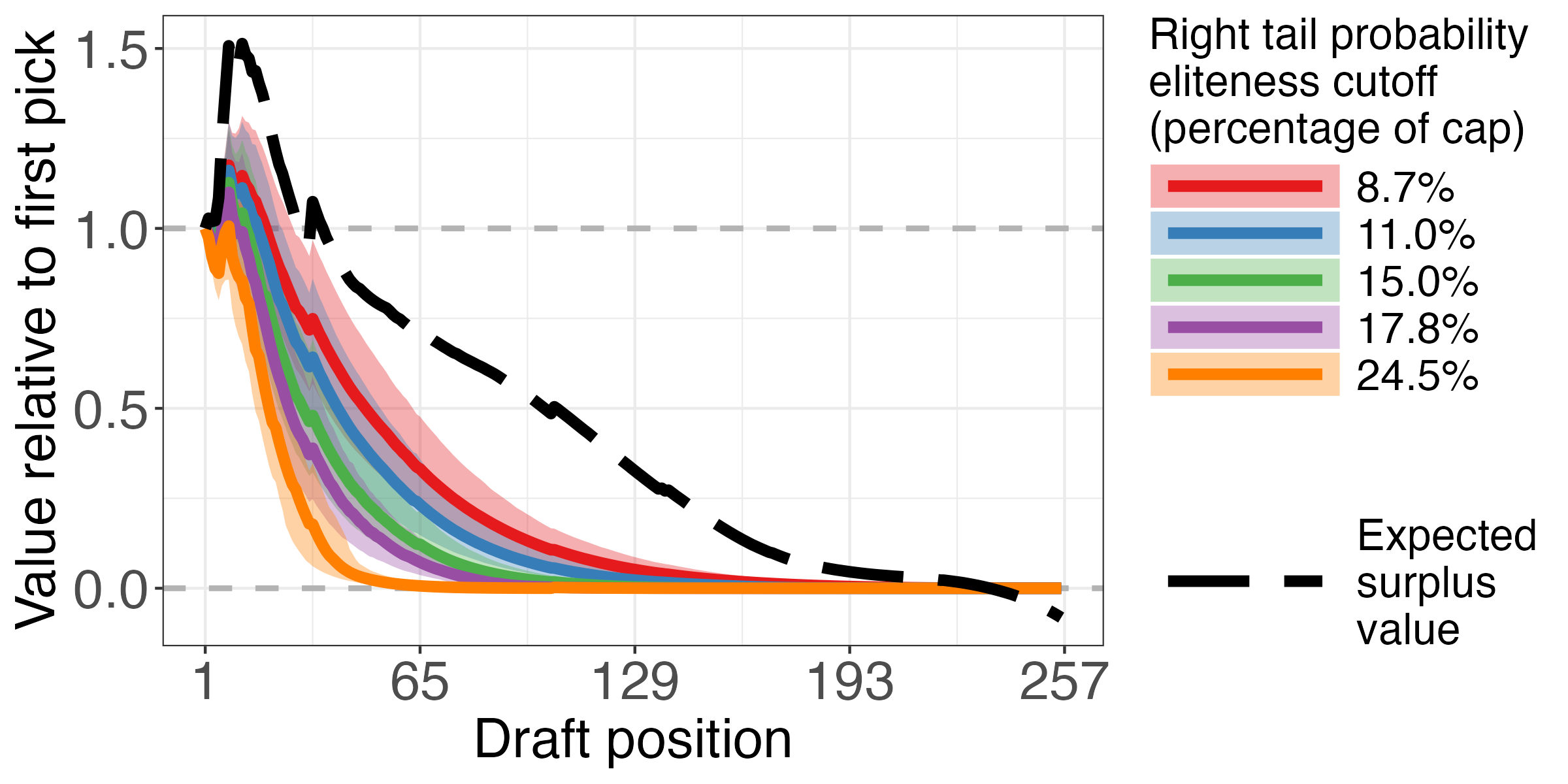}
    \caption{
        Value relative to the first pick ($y$-axis) as a function of draft position $x$. 
        The long-dashed black line is the expected surplus value curve.
        The colored lines are proportional to right tail probability $x \mapsto \bP(S>r|x)$, where $r$ is an eliteness cutoff (color).
        The lines are the posterior means and the shaded regions are the $95\%$ credible intervals.
    }
    \label{fig:plot_G_surplusValueCurves}
\end{figure}

In Figure~\ref{fig:plot_G_surplusValueCurves} we visualize expected surplus value (black dashed line) and the probability that surplus value exceeds a threshold (colored lines).
Massey and Thaler's loser's curse––top draft picks yield lower expected surplus value than those selected later in the first round––manifests as a spike in the expected surplus value curve above one during the first round.

Although the point estimate of the probability that surplus value exceeds a high threshold has a spike––though one that is smaller than and earlier in the draft than that of expected surplus value––consideration of the entire posterior distribution shows that a monotonically decreasing relationship is highly compatible with the data and model.

%%%%%%%%%%%%%%%%%%
%%%%%%%%%%%%%%%%%%

%%%%%%%%%%%%%%%%%%%%%%%%%%%%%%%%%%%%%%%%%%%%%%%%%%%%%%%%%%%%%%%%%%%%%%%%%%%%%%%%%%%%%%%%%
\subsection{Adjusting for position}\label{sec:pos}

The distribution of performance value varies not only by draft position but also by player position (see Section~\ref{app:estimate_con_density_pos}). 
Quarterback performance in particular exhibits a substantially fatter right tail than other positions (see Figure~\ref{fig:plot_post_density_full_byQB}). 
Hence, in this section, we construct separate draft position value curves for quarterbacks and non-quarterbacks.

\begin{figure}[hbt!]
    \centering{}
    \includegraphics[width=\textwidth]{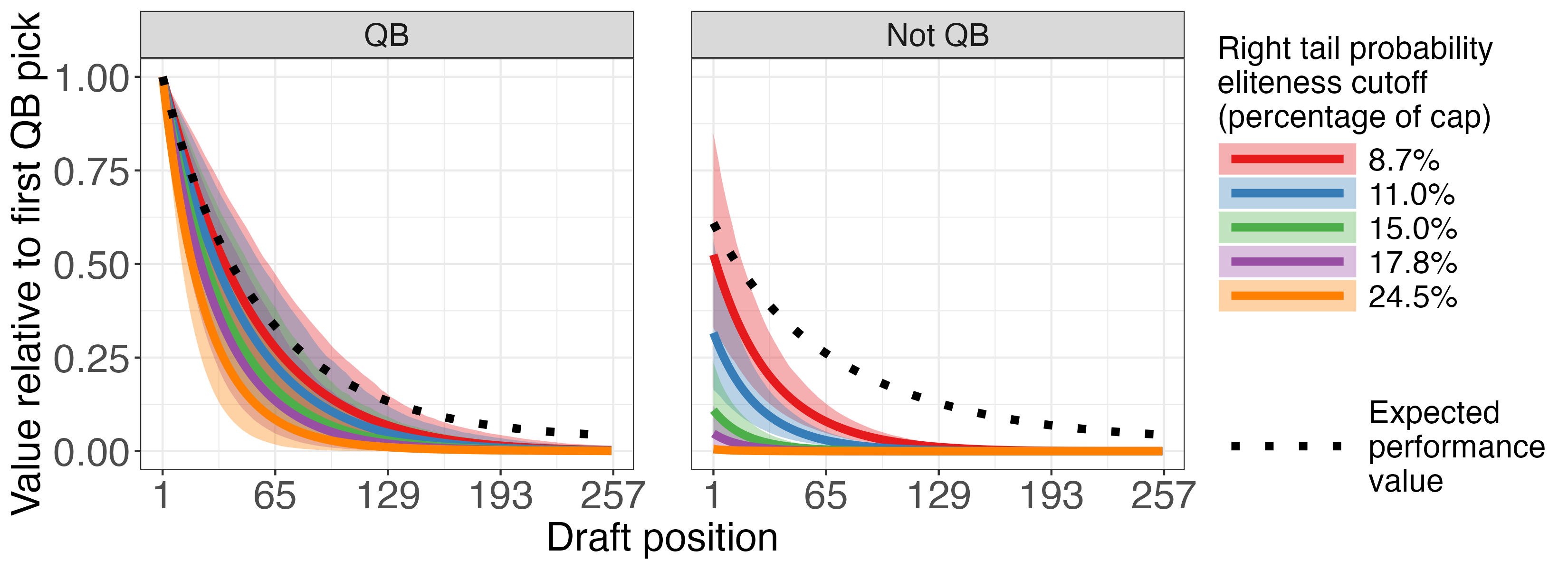}
    \caption{
        Value relative to the first $\qb$ pick ($y$-axis) as a function of draft position $x$ and position (facet).  
        The dotted black line is expected performance value.
        The colored lines are proportional to right tail probability $x \mapsto \bP(Y>r|x,\pos)$, where $r$ is an eliteness cutoff (color).
        The lines are posterior means and the shaded regions are $95\%$ credible intervals.
    }
    \label{fig:plot_G_valueCurves_byQB}
\end{figure}

We begin by visualizing performance value curves in Figure~\ref{fig:plot_G_valueCurves_byQB}.
For $\pos \in \{\qb, \notqb\}$, we compute expected performance value (proportional to $\bE[Y|x,\pos]$) and right tail probability (proportional to $\bP(Y>r|x,\pos)$) from the estimated conditional density $\bP(Y|x,\pos)$.
We compute value relative to a first pick $\qb$.
The position-specific curves exhibit the same broad pattern observed in the overall draft position value curves from Figure~\ref{fig:plot_tail_probs}: right tail probability curves are steeper than expected performance value curves. 
As the eliteness threshold $r$ increases, right tail probability curves become progressively steeper, reinforcing the idea that teams prioritizing increasingly elite talent assign greater relative value to earlier picks.

%%%%%%%%%%%%%%%%%%
%%%%%%%%%%%%%%%%%%

\begin{figure}[hbt!]
    \centering{}
    \includegraphics[width=\textwidth]{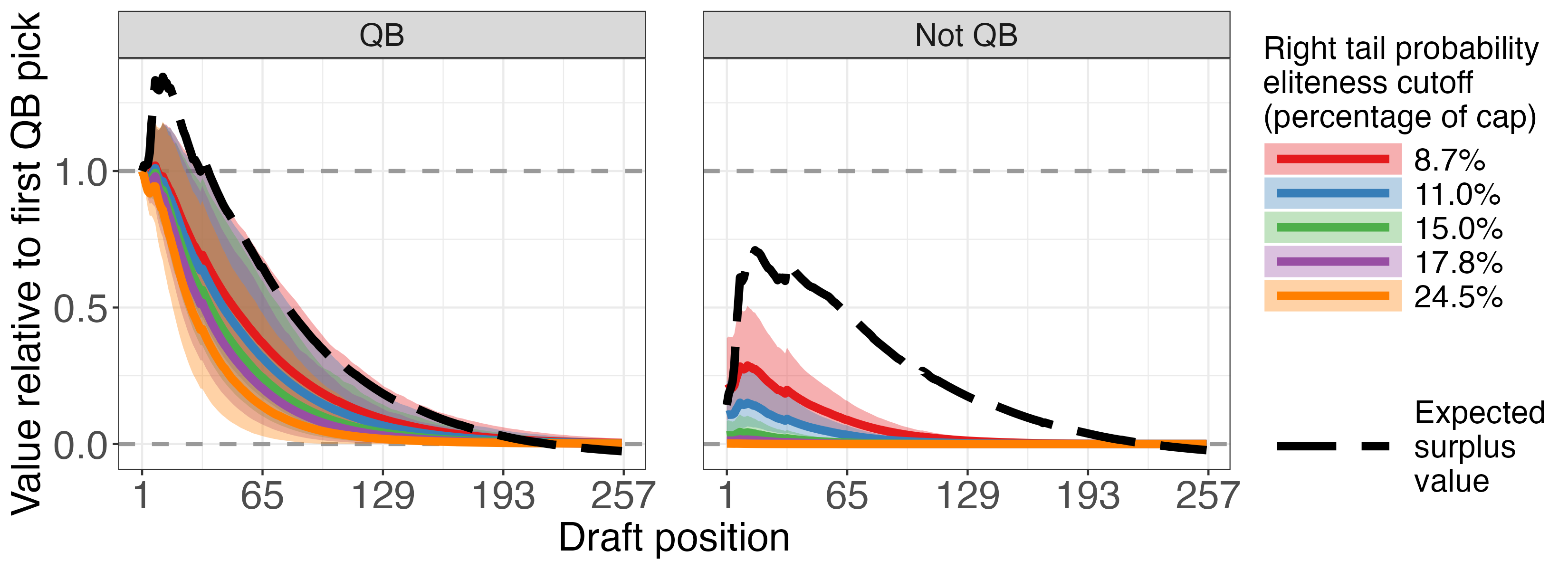}
    \caption{
        Value relative to the first $\qb$ pick ($y$-axis) as a function of draft position $x$ and position (facet).  
        The long-dashed black line is expected surplus value.
        The colored lines are proportional to right tail surplus probability $x \mapsto \bP(S>r|x,\pos)$, where $r$ is an eliteness cutoff (color).
        The lines are posterior means and the shaded regions are $95\%$ credible intervals.
    }
    \label{fig:plot_G_surplusValueCurves_byQB}
\end{figure}

In Figure~\ref{fig:plot_G_surplusValueCurves_byQB} we visualize position-specific surplus value curves.
For both quarterbacks and non-quarterbacks, expected surplus value peaks in the first round.

Although the point estimate of the probability that surplus value exceeds a high threshold has a spike for non-quarterbacks, there is no spike for quarterbacks. 
Further, consideration of the entire posterior distribution for both quarterbacks and non-quarterbacks shows that a monotonically decreasing relationship is highly compatible with the data and model.
Thus, particularly for teams seeking elite quarterbacks, our analysis provides strong support for the notion that the extreme upside of top selections offsets their higher cost.

%%%%%%%%%%%%%%%%%%
%%%%%%%%%%%%%%%%%%

%%%%%%%%%%%%%%%%%%%%%%%%%%%%%%%%%%%%%%%%%%%%%%%%%%%%%%%%%%%%%%%%%%%%
%%%%%%%%%%%%%%%%%%%%%%%%%%%%%%%%%%%%%%%%%%%%%%%%%%%%%%%%%%%%%%%%%%%%
\section{Discussion}\label{sec:discussion}

A central premise in the judgment and decision making literature is that experts, even in high-stakes environments, exhibit systematic cognitive biases. \citet{MasseyThaler2013} applied this framework to the NFL draft, asserting that, according to expected surplus value, general managers overvalue top draft picks, a finding widely cited as evidence of irrational decision making. This assertion, however, hinges on the assumption that general managers should optimize for expected surplus value, an assumption that is neither explicitly justified nor necessarily aligned with the strategic realities of NFL team-building.

In this paper, we challenge that assumption and present an alternative framework for evaluating the value of a draft position. 
By estimating the full conditional density of player performance given draft position and player position, we construct draft position value curves from alternative utility functions.
We focus on right tail probability––equivalently, the number of elite players expected to arise from a given pick––reflecting that general managers may be prioritizing the acquisition of elite, game-changing players. 
Under this perspective, top draft picks are not inefficiently overvalued but rather priced for their potential to produce transformational talent—players who can dramatically elevate a franchise's trajectory.
When valuing draft position based on right tail probability rather than expected surplus value, the first-round spike in the value of a draft position largely disappears. 
Teams selecting at the top of the draft retain higher relative value when their objective is to secure elite talent rather than merely optimizing expected surplus value.
% expected returns. 
This effect is particularly pronounced for quarterbacks, whose performance distributions exhibit extremely fat right tails. 

This study contributes to the judgment and decision making literature by highlighting the importance of specifying the decision-maker’s utility function before labeling behavior as irrational. If general managers value variance and right-tail outcomes rather than expected surplus value, then their draft strategies, far from being biased or inefficient, may be consistent with a rational objective function tailored to the realities of NFL competition. 
Perhaps irrationality and biases are not as pervasive as Massey and Thaler, along with the broader judgement and decision making literature, assert. More broadly, our findings underscore the risks of applying oversimplified economic models to complex, domain-specific decision-making environments without considering alternative utility functions.

While our analysis provides a compelling alternative to the traditional expected performance value and expected surplus value framework, it does not resolve the question of how general managers ought to value draft picks. 
A more comprehensive analysis would explicitly estimate the impact of draft selections on Super Bowl win probability, accounting for factors such as team composition––including projected current and future talent at each position––injury risk, salary cap constraints, and projected outcomes of available draft prospects.
Prioritizing expected surplus value may be misguided, but so may prioritizing eliteness and variance; determining the optimal balance remains an open question. 
We suspect the best objective combines elements of both.  

A broader implication of this study is the potential disconnect between formal decision models and the intuitive expertise of practitioners. 
NFL general managers may not be explicitly solving optimization problems or consciously weighing right-tail probabilities, yet their actions suggest an implicit alignment with valuing elite performance.
This raises an important question: are expert decision-makers in other fields adept at intuitively navigating their domains in ways that are difficult for modelers to formalize? 
If so, then economists, statisticians, and decision scientists should approach expert behavior with greater humility before declaring it irrational. 
While analytical models provide valuable insights, they may overlook nuanced decision making processes that emerge from experience, tacit knowledge, and the complexity of real-world environments.

At the same time, our assumption that general managers' primary goal is to win the Super Bowl may not be entirely correct. It is possible that NFL general managers optimize for other outcomes, such as improving their win-loss record, making the playoffs, or simply preserving their own job security. Moreover, even if their decisions align with maximizing the probability of drafting elite players, this does not necessarily imply rationality. Their choices could stem from overconfidence in their ability to identify talent, a misperception of the distribution of player outcomes, or other psychological biases. The fact that the trade market aligns with a $19.7\%$ of salary cap threshold of eliteness may be a coincidence rather than the product of deliberate strategic thinking. Future research should explore how expert intuition interacts with formal modeling across domains, identifying where models enhance decision making and where they risk oversimplifying the expertise of practitioners.

Finally, we conclude with a discussion of technical limitations of our study and directions for future work.
First, the methodology underlying the fitted trade market curve (the Weibull curve from Appendix~\ref{app:trade_market_value}) has imperfections, which we detail in that section.
In particular, if the team trading down uses a different value function than the team trading up, how do we explain trading? 
Second, in future work, we suggest exploring curves built from the sum of performance and surplus, rather than each one individually. 
When drafting a player, a general manager benefits from the value of his performance in addition to the extra cash from a cheap rookie contract.
Third, a major assumption employed in this work and in previous work is additivity: the value of a bundle of draft picks $v(\{x_j\})$ equals its sum, $\sum_j v(x_j)$. 
Concretely, in terms of right tail probability, we consider the value of a bundle of picks by $\sum_j \bP(Y_j > r)$, though the relationship between the picks may be non-additive or something more general, like $\bP(f(Y_1,...,Y_n) > r)$.
We leave an exploration of non-additve draft position value curves to future work.
Fourth, in this study we consider just first contract performance and surplus value. 
First contract value is an incomplete measure of the total value he provides to his team, as it excludes the rest of his career.
Finally, we did not address a selection bias that was explored in \citet{NFLDradftSelectionBias}: worse teams draft players earlier in each round of the draft than better teams, and a team's quality may influence the ultimate performance outcome of a drafted player.
A more elaborate construction of NFL draft trade value curves should account for this selection bias, which we leave to future work.

% %%%% REFERENCES
% \clearpage
% \newpage
% \bibliography{refs}
% \end{document}

\if0\blind
{
  \section*{Acknowledgments}
  The authors thank Cade Massey, Christopher Avery, and Blake Zilberman for giving the authors a reason to explore draft value curves.
We also thank Cade and Chris for sending NFL draft data for an early version of this project.
} \fi

\bibliography{refs}
% \bibliography{refs.bib}
\newpage
% \bigskip
\begin{center}
{\large\bf SUPPLEMENTARY MATERIAL}
\end{center}
\appendix
% %%%%%%%%%%%%%%%%%%%%%%%%%%%%%
% \documentclass[12pt]{article}
% \usepackage{amsmath,amsfonts,amssymb}
% \usepackage{graphicx,psfrag,epsf}
% \usepackage{enumerate}
% \usepackage[round]{natbib}
% \usepackage{url} 
% \usepackage[dvipsnames]{xcolor}
% \input{header}
% % \doublespace
% \onehalfspacing
% \bibliographystyle{apalike}
% % DON'T change margins - should be 1 inch all around.
% \addtolength{\oddsidemargin}{-.5in}%
% \addtolength{\evensidemargin}{-.5in}%
% \addtolength{\textwidth}{1in}%
% \addtolength{\textheight}{1.3in}%
% \addtolength{\topmargin}{-.8in}%

% \usepackage{fullpage}
% \usepackage{parskip}
% % \usepackage{mathptmx}
% % \newtheorem{theorem}{Theorem}
% % \numberwithin{equation}{section}
% \interfootnotelinepenalty=10000
% \definecolor{SkyBlue}{RGB}{14, 118, 188}
% \definecolor{BrightRed}{RGB}{223,82, 78}
% \hypersetup{pdfborder = {0 0 0.5 [3 3]}, colorlinks = true, linkcolor = BrightRed, citecolor = SkyBlue, filecolor = BrightRed}
% %%%%%%%%%%%%%%%%%%%%%%%%%%%%%

% \begin{document}

% \begin{center}
% {\large\bf SUPPLEMENTARY MATERIAL}
% \end{center}
% \appendix

%%%%%%%%%%%%%%%%%%%%%%%%%%%%%%%%%%%%%%%%%%%%%%%%%%%%%%%%%%%%%%%%%%%%
%%%%%%%%%%%%%%%%%%%%%%%%%%%%%%%%%%%%%%%%%%%%%%%%%%%%%%%%%%%%%%%%%%%%
%%%%%%%%%%%%%%%%%%%%%%%%%%%%%%%%%%%%%%%%%%%%%%%%%%%%%%%%%%%%%%%%%%%%
\section{Our code}\label{app:code}

Our analysis in this paper is reproducible via the code that is publicly available on Github at 
\if0\blind
{
  \url{https://github.com/snoopryan123/NFL_draft_chart_Ryan}.
} \fi
\if1\blind
{
  [URL blinded for review].
} \fi

%%%%%%%%%%%%%%%%%%%%%%%%%%%%%%%%%%%%%%%%%%%%%%%%%%%%%%%%%%%%%%%%%%%%%%%%%%%%%%%%%%
\section{Measuring a player's performance value}\label{app:performanceValue}

Each play in American football consists of a complex interplay between 22 players of varying roles interacting in continuous time and space, making it difficult to measure how a player's individual performance impacts the ultimate outcome of the play.
For instance, a linebacker may impact the outcome of a play simply by occupying space, causing the ball carrier to move in a different direction, and there is no easy way to measure the impact of that action.
Since there is no easy or industry-standard way of measuring player performance on a given play, particularly for off-the-ball players, analysts use simpler proxies of performance value.
Importantly, these proxies measure the performance of players of all positions on the same scale.
Each of these are imperfect measures with their own drawbacks.

In free agency, NFL players who are not bound by a contract can sign a new contract with any team. %, and the value of this contract is bounded only by the salary cap.
Assuming the free agent market is efficient, it is a reflection of the true performance value of players.
Hence, one popular measure of performance value is inflation-adjusted salary. %in dollars.
\citet{BaldwinPosCurves} adjusts for inflation by measuring salary as a percentage of the salary cap, referred to as APY cap percentage (APY means average salary per year).
\citet{OTCDraftChart} measure a player's performance value as his salary as a percentage of the average of the top five salaries at his position.
For example, in 2024 the top five quarterback salaries have an APY of \$51.2 million, so they grade Mahomes' APY of \$45 million as an 88\%.
A drawback of measuring a player's value relative to other players at his position is it doesn't account for positional value, or the fact that some positions are inherently more valuable than others (e.g., an average quarterback is much more valuable than an average running back).
A drawback of using the free agent market to measure performance is that the market is a reflection of projected future value, not past performance value.
A running back who performs very well in his first contract may not receive a large second contract because running backs are injury prone and age quickly.
For example, Saquon Barkley, one of the best running backs in the NFL during his rookie contract, signed a modest second contract in 2023 worth \$10.1 million for one year.
His free agent contract surely understates his first contract performance value.

To measure past performance value rather than projected future performance value, \citet{MasseyThaler2013} use the number of games a player started and whether he made the Pro Bowl.
They then map this performance to salary by regressing a player's second contract compensation (the value of a player's first free agent contract) on his number of starts and Pro Bowls in his first contract (see Section 6.2 of their paper for the particulars of this regression).
A drawback of their approach is that the number of starts and Pro Bowls is a crude measure of player performance.
Two linebackers can have the same number of starts and Pro Bowls but have vastly different impacts on the games they played, for instance by having a different number of tackles or yards allowed.

\citet{StuartDraftChart} measures player performance using Pro Football Reference's Approximate Value (AV) \citep{PfrAV}.
AV maps more granular position-specific performance measures to a common scale.
For offensive players, AV apportions the points per drive scored by a team's offense to each of its offensive players in proportion to position-specific performance measures.
For skill position players, AV uses yards to measure performance (e.g., rushing yards for running backs and pasing yards for quarterbacks).
AV uses more crude measures for offensive lineman (e.g., games played, games started, and All Pro).
AV divides credit for an offensive team's points scored per drive using \textit{ad hoc} multipliers (e.g., the offensive line gets $5/11$ of the offensive credit since there are usually 5 offensive lineman on the field).
AV works analogously for defensive players.
Drawbacks of AV are that it is too simple, too \textit{ad hoc}, and uses crude performance measures for non-skill position players.

\citet{PFFDraftChart} (known as PFF) measures player performance using Wins Above Replacement (WAR) \citep{eager2020}.
WAR estimates an individual player’s contribution to team performance on the scale of team wins relative to a replacement-level player (e.g., see \citet{gridWar} for an exposition of WAR for starting pitchers in baseball).
PFF grades the performance of every player in every football game from -2 to +2 in increments of $1/2$, where 0 represents average performance.
They then estimate WAR from these PFF grades using a Massey rating system \citep{Massey97}.
A drawback of PFF WAR is that it is a model-based metric and it is unclear how good or reliable this model is at measuring player performance.
Some people take issue with PFF grades, which are subjective measures of player performance made by humans who watch the tape.
For instance, a Pro Bowl NFL offensive lineman once told us that PFF often grades him incorrectly because it doesn't know the job his coach told him to do.
He should've received a perfect grade for doing his job, which is not publicly available knowledge, even though watching the tape may make it seem like he played suboptimally.

%%%%%%%%%%%%%%%%%%%%%%%%%%%%%%%%%%%%%%%%%%%%%%%%%%%%%%%%%%%%%%%%%%%%
%%%%%%%%%%%%%%%%%%%%%%%%%%%%%%%%%%%%%%%%%%%%%%%%%%%%%%%%%%%%%%%%%%%%
%%%%%%%%%%%%%%%%%%%%%%%%%%%%%%%%%%%%%%%%%%%%%%%%%%%%%%%%%%%%%%%%%%%%
\section{Estimating the trade market value of draft picks}\label{app:trade_market_value}

We follow the approach from \citet{MasseyThaler2013} to estimate the relative value of draft picks that is implied by the NFL trade market.
This trade market value function $\vmarket$ is a function of a bundle of picks $\{x_j\}_{j=1}^{n}$, where each pick $x_j$ belongs to $\{1,...,256\}$.
Without loss of generality, assume that picks in a bundle are ordered from earliest to latest pick, $x_1 \leq ... \leq x_n$.
Also, $\vmarket(\{1\})=\vmarket(1)=1$ so that value is relative to the first pick.

\citet{MasseyThaler2013} make simplifying assumptions in fitting $\vmarket$.
First, they assume additivity,
\begin{equation}
\vmarket(\{x_j\}_{j=1}^{n}) = \sum_{j=1}^{n} \vmarket(x_j).
\label{eqn:market_value_additive}
\end{equation}
Second, they suppose that all teams use the same market value function $\vmarket$ to value draft picks. %\fixme{on average}.
Third, they assume that trades in the market are fair, so the market value of both teams' bundles of picks are equal. 
% are fair \fixme{on average}
Formally, denoting the bundle of picks initially belonging to the team trading down by $\{x_j^{(d)}\}_{j=1}^{n_d}$ and that of the team trading up by $\{x_j^{(u)}\}_{j=1}^{n_u}$, the trade is fair if 
% $\sum_{j=1}^{n_d}\vmarket(\{x_j^{(d)}\}) = \sum_{j=1}^{n_u}\vmarket(\{x_j^{(u)}\})$.
\begin{equation}
\sum_{j=1}^{n_d}\vmarket(\{x_j^{(d)}\}) = \sum_{j=1}^{n_u}\vmarket(\{x_j^{(u)}\}).
\label{eqn:market_value_fair}
\end{equation}
Finally, they assume the market value of picks follows a Weibull distribution, 
% $\vmarket(x) = e^{-\lambda(x-1)^\beta},$ 
\begin{equation}
\vmarket(x) = e^{-\lambda(x-1)^\beta},
\label{eqn:market_value_weibull}
\end{equation}
where $\lambda$ and $\beta$ are parameters to be estimated.
They also consider a Weibull model with a discount rate $\rho$ for a pick $N$ years into the future, 
% $\vmarket(x,N) = e^{-\lambda(x-1)^\beta}\cdot(1+\rho)^N.$
\begin{equation}
\vmarket(x,N) = e^{-\lambda(x-1)^\beta}\cdot(1+\rho)^N.
\label{eqn:market_value_weibull_discount}
\end{equation}
Combining equations \eqref{eqn:market_value_fair}, \eqref{eqn:market_value_additive}, and \eqref{eqn:market_value_weibull}, they calculate
\begin{equation}
\log(x^{(d)}_1) = \log\bigg( \bigg[ -\frac{1}{\lambda}\log\bigg( \sum_{j=1}^{n_u} e^{-\lambda(x_j^{(u)}-1)^\beta} - \sum_{j=2}^{n_d} e^{-\lambda(x_j^{(d)}-1)^\beta} \bigg) \bigg]^{1/\beta} \bigg),
\label{eqn:weibull_nonlinear_reg}
\end{equation}
which expresses the value of the top pick in the trade acquired by the team trading down in terms of the other picks in the trade. 
They estimate the parameters $\lambda$ and $\beta$ in Equation~\eqref{eqn:weibull_nonlinear_reg} using nonlinear regression.
They use a similar approach to find a trade market value curve that incorporates the discount rate.

The aforementioned assumptions range from quite strong to very strong and deserve to be examined.
The fourth assumption, that $\vmarket$ follows a Weibull distribution, seems to be the most mild, as the Weibull is a sufficiently expressive family that can capture many different plausible draft position value curves.
The first assumption, additivity, is wrong but mild. 
For instance, the value of a bundle of $12$ picks is not additive because not all of these players can play on the field at the same time.
But additivity doesn't seem to be a terrible assumption because most trades involve just a few picks.
The third assumption is wrong; they probably meant to assume that observed trades are fair \textit{on average}:
\begin{equation}
\sum_{j=1}^{n_d}\vmarket(\{x_j^{(d)}\}) = \sum_{j=1}^{n_u}\vmarket(\{x_j^{(u)}\}) + \varepsilon,
\label{eqn:market_value_fair_onAvg}
\end{equation}
where $\varepsilon$ is mean zero noise.
The analogue of Equation~\ref{eqn:weibull_nonlinear_reg} under this new assumption is
\begin{equation}
\log(x^{(d)}_1) = \log\bigg( \bigg[ -\frac{1}{\lambda}\log\bigg( \sum_{j=1}^{n_u} e^{-\lambda(x_j^{(u)}-1)^\beta} - \sum_{j=2}^{n_d} e^{-\lambda(x_j^{(d)}-1)^\beta}  + \varepsilon \bigg) \bigg]^{1/\beta} \bigg),
\label{eqn:weibull_nonlinear_reg_eps}
\end{equation}
which is not equal to 
\begin{equation}
\log(x^{(d)}_1) = \log\bigg( \bigg[ -\frac{1}{\lambda}\log\bigg( \sum_{j=1}^{n_u} e^{-\lambda(x_j^{(u)}-1)^\beta} - \sum_{j=2}^{n_d} e^{-\lambda(x_j^{(d)}-1)^\beta} \bigg) \bigg]^{1/\beta} \bigg) + \varepsilon.
\label{eqn:weibull_nonlinear_reg_eps2}
\end{equation}
The nonlinear regression they employ implicitly uses the form from Equation~\eqref{eqn:weibull_nonlinear_reg_eps2} even though they should fit from Equation~\eqref{eqn:weibull_nonlinear_reg_eps}.
Notably, however, the second assumption (that all teams use the same market value function $\vmarket$ to value draft picks) is the most egregious.
Each general manager has his own value function, which changes from year to year and can be skewed by meddling owners or coaches.
We could potentially treat the value function employed in each trade as a draw centered at $\vmarket$, with multiple levels of variance for each general manager and each year.
There is likely not enough data to estimate such a hierarchical model well and its standard error would be large.
 % but there is nowhere near enough data to  and the standard errors would blow up, leaving us with no insight on the market.
% If we treat the value function employed in each trade as a draw centered at $\vmarket$, then $\varepsilon$ absorbs the additional noise from this draw.
% On this view, we think of $\vmarket$ as an ``average'' market value curve.

\begin{figure}[hbt!]
    \centering{}
    \includegraphics[width=\textwidth]{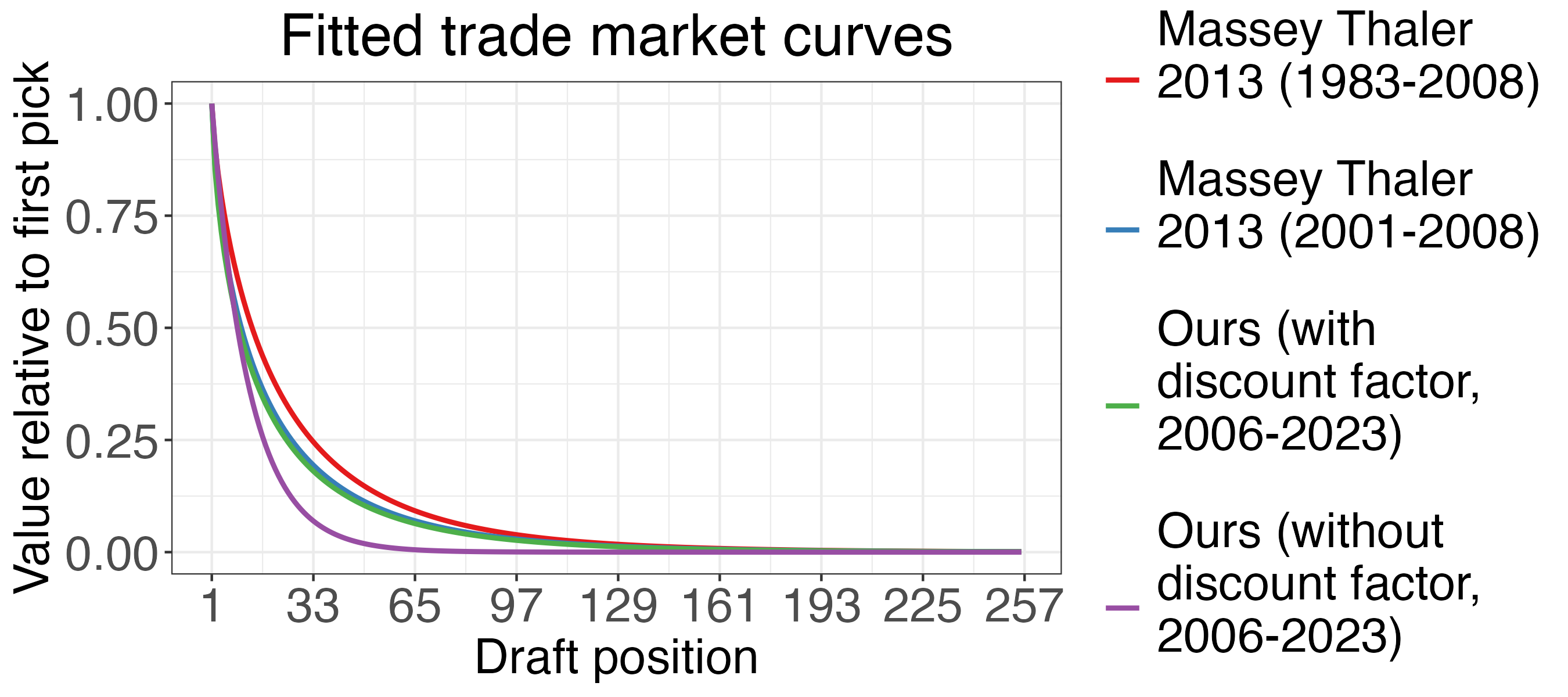}
    \caption{
        Weibull trade market value curves $\vmarket$ according to \citet{MasseyThaler2013} and our replication.
        Massey and Thaler's curves here do not use a discount factor. 
    }
    \label{fig:plot_market_curves}
\end{figure}

Despite these issues with the simplifying assumptions, we replicate the analysis discussed above from a dataset consisting of all draft trades consisting purely of draft picks from 2006-2023.
%%%%Restricting the dataset to recent trades from 2013-2018 hardly changes the shapes of the curves.
In light of the aforementioned flaws, we use the fitted $\vmarket$ curve to tell a general story about the steepness of the trade market.
We take it with a grain of salt without obssessing over the exact shape of the curve or its minutae.
We display the $\vmarket$ from \citet{MasseyThaler2013} and from our replication, with and without the discount factor, in Figure~\ref{fig:plot_market_curves}.
%%%%%
These curves share the same general shape––a monotonically decreasing and extremely steep trend––indicating general managers' stark preference for top picks.
Value declines rapidly, falling to between one-half and one-quarter of a first pick by the middle of the first round.  
Value then decays to less than a tenth of a first pick by the start of the second round before quickly plummeting to near-zero thereafter.

%%%%%%%%%%%%%%%%%%
%%%%%%%%%%%%%%%%%%

% We find a discount factor of $\rho = -0.584$, indicating that the market finds a pick next year about $40\%$ less valuable than if it were this year.
% The estimated Weibull parameters for $\rho=0$ are $(\beta=0.975, \lambda=0.091)$ and when we include a discount factor are $(\beta=0.686, \lambda=0.159, \rho=-0.584)$.
% Bootstrapping, which doesn't account for all the sources of variation discussed in the previous paragraph but is informative nonetheless, reveals large standard errors in the Weibull parameters and the discount factor (not shown).

% \bibliography{refs}
% \end{document}

\end{document}